\documentclass[conference]{IEEEtran}

\newif\ifapx
\apxtrue

\newif\ifblind
\blindfalse

\newif\ifmulticlass
\multiclasstrue

\newif\ifkeywords
\keywordstrue
\newif\ifacknowledge
\acknowledgefalse

\newcommand{\ourmaintitle}{Efficiently Discovering Locally Exceptional yet Globally Representative Subgroups}

\usepackage{latexsym}
\usepackage[ruled, vlined, nofillcomment, linesnumbered]{algorithm2e}
\usepackage[noenumitem]{eda}

\graphicspath{ {./figs/} }

\ifpdf
\usetikzlibrary{external}
\fi

\newcommand{\tvar}[1]{\texttt{#1}}
\newcommand{\tdef}[1]{\textbf{#1}}
\newcommand{\tDS}[1]{\emph{#1}}
\newcommand{\tpred}[1]{\texttt{#1}}
\newcommand{\temph}[1]{\textbf{#1}}
\newcommand{\tmale}{\mars}
\newcommand{\tfemale}{\venus}
\newcommand{\tyes}{\cmark}
\newcommand{\tno}{\xmark}
\newcommand{\ifmce}[2]{%
\ifmulticlass%
#2%
\else%
#1%
\fi%
}
\newcommand{\ifmc}[2][]{\ifmce{#1}{#2}}
\newcommand{\proofat}[1]{(for the proof see Appendx~\ref{#1})}
\newcommand{\mathnomath}[1]{\ensuremath{#1}\ifmmode\else\xspace\fi}
\newcommand{\eqdef}{\coloneqq}

\newcommand{\fVec}[1]{\mathnomath{{\boldsymbol{#1}}}}

\newcommand{\ifeab}[4][]{\ifthenelse{\equal{#1}{#2}}{#3}{#4}}
\newcommand{\ifne}[3][]{\ifthenelse{\equal{#2}{#1}}{}{#3}}

\newcommand{\sfcard}[1]{|#1|}

\newcommand{\sitem}{\mathnomath{e}}
\newcommand{\sritem}{\mathnomath{\sitem}}

\newcommand{\sVars}{\mathnomath{{\fVec{V}}}}

\newcommand{\sLang}{\mathnomath{\mathcal{L}}}
\newcommand{\sLMem}[1][]{\mathnomath{\sigma\ifne{#1}{_{#1}}}}
\newcommand{\sLRef}{\mathnomath{\rho}}

\newcommand{\sfext}{\mathnomath{\operatorname{ext}}}
\newcommand{\sffit}{\mathnomath{f}}
\newcommand{\sfFitCCS}[3][]{\mathnomath{\sffit\ifsub{#1}^{\mathsmaller{#2}} \ifne{#3}{(#3)}}}
\newcommand{\sffitCCS}[2][\sSub]{\sfFitCCS[]{#1}{#2}}
\newcommand{\sReprCCS}[2][\sSub]{\sfFitCCS[\textup{r}]{#1}{#2}}
\newcommand{\sCovCentCCS}[2][\sSub]{\sfFitCCS[\textup{ct}]{#1}{#2}}
\newcommand{\sCentCCS}[2][\sSub]{\sfFitCCS[\textup{t}]{#1}{#2}}
\newcommand{\sfoest}{\mathnomath{\hat\sffit}}

\newcommand{\sPop}{\mathnomath{P}}
\newcommand{\svctr}{\mathnomath{c}}
\newcommand{\svtar}{\mathnomath{y}}

\newcommand{\sSub}{\mathnomath{Q}}
\newcommand{\sSel}{\mathnomath{R}}

\newcommand{\sCov}{\mathnomath{\sffit_{\textup{c}}}}

\newcommand{\sCent}{\mathnomath{\sffit_{\textup{t}}}}
\newcommand{\sRepr}{\mathnomath{\sffit_{\textup{r}}}}
\newcommand{\sCovCent}{\mathnomath{\sffit_{\textup{ct}}}}
\newcommand{\sProb}{\mathnomath{\mathbb{P}}}
\newcommand{\sfdist}[1][]{\mathnomath{d\ifne{#1}{_{\textup{#1}}}}}
\newcommand{\sftvd}{\sfdist[TV]}

\newcommand{\ifsub}[1]{\ifthenelse{\equal{#1}{}}{}{_{#1}}}
\newcommand{\ifnsv}[2]{\ifthenelse{\equal{#2}{}}{\fVec{#1}}{\ifthenelse{\equal{#2}{*}}{#1}{{#1}_{#2}}}}
\newcommand{\fnSet}[2][\siCls]{\mathnomath{\ifnsv{n}{#1} (#2)}}
\newcommand{\snPop}[1][]{\mathnomath{\ifnsv{n}{#1}}}
\newcommand{\snSub}[1][]{\mathnomath{\ifnsv{m}{#1}}}

\newcommand{\snCls}{\mathnomath{K}}
\newcommand{\siCls}{\mathnomath{k}}
\newcommand{\fpSet}[2][\siCls]{\mathnomath{\ifnsv{p}{#1} (#2)}}
\newcommand{\spPop}[1][]{\mathnomath{\ifnsv{p}{#1}}}
\newcommand{\spSub}[1][]{\mathnomath{\ifnsv{q}{#1}}}
\newcommand{\spSel}[1][]{\mathnomath{\ifnsv{r}{#1}}}

\newcommand{\sPInd}[1][]{\mathnomath{\ifnsv{I}{#1}}}
\newcommand{\sSelPar}[1][\sPInd]{\mathnomath{\mathcal{R}_{#1}}}
\newcommand{\sSelParOpt}[1][\sPInd]{\mathnomath{R^*_{#1}}}
\newcommand{\sPIndSet}{\mathnomath{\mathcal{I}}}
\newcommand{\sfvtar}[2][\siCls]{\mathnomath{\svtar\ifsub{#2}\ifeab{#1}{}{^{(#1)}}}}
\newcommand{\sfvctr}[1]{\mathnomath{\svctr\ifsub{#1}}}

\newcommand{\siSub}{\mathnomath{i}}
\newcommand{\siCnt}{\mathnomath{\mu}}
\newcommand{\siSeq}{\mathnomath{\tau}}
\newcommand{\scSeq}{\mathnomath{t}}
\newcommand{\sDenom}[2][]{\mathnomath{\nu_{\textup{#2}\ifne{#1}{(#1)}}}}
\newcommand{\sSBVec}[1][\siCls]{\mathnomath{\fVec{e}_{#1}}}

\newcommand{\sconstt}{\mathnomath{\alpha_{\textup{t}}}}
\newcommand{\sconstc}{\mathnomath{\alpha_{\textup{c}}}}
\newcommand{\sweight}{\mathnomath{{\gamma}}}

\newcommand{\sReprScale}{\mathnomath{a}}
\newcommand{\sPath}[2][]{\mathnomath{\ifnsv{\pi}{#1}\ifne{#2}{\ifeab[#2]{*}{^{#2}}{^{\mathsmaller(#2\mathsmaller)}}}}}
\newcommand{\sPathOptInd}{\mathnomath{\mu^*}}

\newcommand{\sSeqA}[1][]{\mathnomath{a\ifne{#1}{^{(#1)}}}}
\newcommand{\sSeqDom}{\mathnomath{N}}
\newcommand{\snSeq}{\mathnomath{n}}
\newcommand{\sRegion}[1]{\mathnomath{A_{\textup{#1}}}}

\newcommand{\sfmap}[3]{\mathnomath{#1\!:#2\to#3}}

\newcommand{\fReals}[1][]{\mathnomath{\mathbb{R}\ifne{#1}{^{#1}}}}
\newcommand{\fBools}[1][]{\mathnomath{\{\top,\bot\}\ifne{#1}^{#1}}}
\newcommand{\sHSeq}[2][\siSub_2]{\mathnomath{\fVec{h}_{\fVec{#1}}\ifne{#2}{\ifeab[#2]{*}{^{\fVec{#2}}}{^{\mathsmaller{\fVec{(#2)}}}}}}}
\newcommand{\sVSeq}[2][\siSub_1]{\mathnomath{\fVec{v}_{\fVec{#1}}}\ifne{#2}{\ifeab[#2]{*}{^{\fVec{#2}}}{^{\mathsmaller{\fVec{(#2)}}}}}}

\newcommand{\fTSearch}[3]{\TSearch(\text{on }#1\text{ from }#2\text{ to }#3)}
\newcommand{\sfiAlg}[1]{\mathnomath{\siSub\ifne{#1}{_{\textup{#1}}}}}

\newcommand{\siAlgBeg}{\sfiAlg{beg}}
\newcommand{\siAlgEnd}{\sfiAlg{end}}
\newcommand{\siAlg}{{\sfiAlg{}}}

\newcommand{\sSST}[1][\sSub]{\mathnomath{\mathcal{T}\ifne{#1}{(\sSub)}}}

\newcommand{\sappr}{\mathnomath{\alpha}}
\newcommand{\topt}[1]{\temph{#1}}
\newcommand{\tnopt}[1]{#1}
\newcommand{\tmis}{\ensuremath{\infty}}

\newcommand{\favg}[2][]{\mathnomath{\bar{#2}\ifthenelse{\equal{#1}{}}{}{_{\mathsmaller #1}}}}

\usepackage{bm}
\usepackage{etex}
\usepackage{subcaption}
\usepackage{bbm}
\usepackage{relsize}
\usepackage{todonotes}
\usepackage{pgfplots}
\usepackage{pgfplotstable}
\usepackage{multirow}
\usepackage{tabu}
\usepackage{tabularx}
\usetikzlibrary{shapes}

\usepackage[utf8x]{inputenc}
\usepackage{mathtools}
\usepackage{acro}
\usepackage{tabu}
\usepackage{xcolor}
\usepackage{wasysym}
\usepackage{pifont}
\newcommand{\cmark}{\ding{51}}%
\newcommand{\xmark}{\ding{55}}%
\usepackage[nocompress]{cite}
\newlength{\figWidth}
\newlength{\figHeight}
\DeclareAcronym{BRIG}{short = \textsc{brig},
	long={Binary Representativeness IGnorant}}
\DeclareAcronym{RAWR}{short = \textsc{rawr},
	long={Representativeness\ AWare\ algoRithm}}
\DeclareAcronym{CCS}{short = \textsc{ccs},
	long={class counting space}}
\DeclareAcronym{SST}{short = SST,
	long={sufficient search triangle}}
\DeclareAcronym{BNB}{short = B\&B,
	long={Branch and Bound}}

\newcommand{\tBRIG}{\ac{BRIG}\xspace}
\newcommand{\tRAWR}{\ac{RAWR}\xspace}
\newcommand{\tBNB}{\ac{BNB}\xspace}
\newcommand{\tSST}{\ac{SST}\xspace}
\newcommand{\tCCS}{\ac{CCS}\xspace}

\SetKwComment{tcpas}{\{}{\}}
\SetCommentSty{textnormal}
\SetArgSty{textnormal}
\SetKw{False}{false}
\SetKw{True}{true}
\SetKw{Null}{null}
\SetKwInOut{Output}{output}
\SetKwInOut{Input}{input}
\SetKw{AND}{and}
\SetKw{OR}{or}
\SetKw{Continue}{continue}
\SetKw{BSearch}{BSearch}
\SetKw{TSearch}{TernarySearch}
\SetKw{Sort}{Sort}

\definecolor{draft janis}{rgb}{0.4,0.4,0.8}
\definecolor{note janis}{rgb}{0.2,0.2,1}

\DeclareMathOperator*{\argmax}{arg\,max}

\DeclareMathOperator{\sign}{sign}

\DeclareMathOperator{\mean}{mean}

\definecolor{ginger}{rgb}{0.69, 0.4, 0.0}
\definecolor{cadmiumred}{rgb}{0.89, 0.0, 0.13}
\definecolor{internationalorange}{rgb}{1.0, 0.31, 0.0}
\definecolor{green(ryb)}{rgb}{0.4, 0.69, 0.2}

\colorlet{cov colour}{ginger}
\colorlet{cent colour}{cadmiumred}
\colorlet{cov colour}{internationalorange}
\colorlet{cent colour}{green(ryb)}

\colorlet{covcent colour}{cov colour!50!cent colour}
\colorlet{repr colour}{electricultramarine}

\def\plotDomOffset{0.3}
\def\plotDomSize{0.3}
\pgfplotsset{
  small ticks/.style={
    xtick style = {},
    ytick style = {},
    every tick label/.append style={
      font=\smaller,
    },
    x label style={
      font=\small,
    },
    y label style={
      font=\small,
    },
  },
  ccs plot 2/.style={
    x axis line style = {thick, normal shift={x}{6pt},},
    y axis line style = {thick, normal shift={y}{6pt},},
    xtick style = {thick,normal shift={x}{6pt},},
    ytick style = {thick,normal shift={y}{6pt},},
    every x tick label/.append style={thick,normal shift={x}{6pt},},
    every y tick label/.append style={thick,normal shift={y}{6pt},},
    x label style={normal shift={x}{5pt},},
    y label style={normal shift={y}{-8pt},},
    xmin=0,
    xmax=10,
    ymin=0,
    ymax=10,
    xtick={0,...,10},
    ytick={0,...,10},
    xtick align=outside,
    ytick align=outside,
    line cap=round,
    clip=false,
    axis lines*=left,
    xlabel={$\fnSet[1]\sSel$},
    ylabel={$\fnSet[2]\sSel$},
  },
  ccs plot/.style={
    xmin=-0.01,
    xmax=10.7,
    xtick={0,...,10},
    ytick={0,...,10},
    xlabel near ticks,
    ylabel near ticks,
    axis lines=middle,
    ymin=-0.01,
    xlabel={$\fnSet[1]\sSel$},
    ylabel={$\fnSet[2]\sSel$},
    xlabel near ticks,
    ylabel near ticks,
    x label style={
      at={(axis description cs:0.5,-0.05)},anchor=north
    },
    y label style={
      at={(axis description cs:-0.05,.5)},
      rotate=0,anchor=south
    },
    ymax=10.7,
    clip mode=individual,
  },
}
\tikzset{
  repr/.style={blue,thick},
  horiz repr/.style={repr,xshift=\domoffx},
  vert repr/.style={repr,yshift=\domoffy},
  covcent/.style={green,thick},
  horiz covcent/.style={covcent,xshift=-\domoffx},
  vert covcent/.style={covcent,yshift=-\domoffy},
  concave/.style={},
  convex/.style={dashed},
  label/.style={fill=white,rounded rectangle,opacity=0.7,text opacity=1},
  point/.style={inner sep=0.05cm},
  sst/.style={mark size=.75ex,mark options={solid,line width=.4ex},
    red,dashed,fill opacity=0.1,fill=red},
  ctpath/.style={green},
  ctcount/.style={ctpath,mark size=1.5ex,mark options={solid}},
  region/.style={ultra thick,red,opacity=0.3},
  concave sequence/.style={red,opacity=0.25,magenta,line width=.3ex},
}  

\pgfkeysifdefined{/pgfplots/table/output empty row/.@cmd}{
  \pgfplotstableset{
    empty header/.style={
      every head row/.style={output empty row},
    }
  }
}{
  \pgfplotstableset{
    empty header/.style={
      typeset cell/.append code={%
        \ifnum\pgfplotstablerow=-1
        \pgfkeyssetvalue{/pgfplots/table/@cell content}{}%
        \fi
      }
    }
  }
}

\ifblind
\author{\IEEEauthorblockN{Submitted for Blind Review}
  \IEEEauthorblockA{~\\~}
}
\else
\author{%
  \IEEEauthorblockN{%
    Janis Kalofolias,
    Mario Boley, and
    Jilles Vreeken
  }
  \IEEEauthorblockA{%
    Max Planck Institute for Informatics and Saarland University\\
    Saarland Informatics Campus, Germany \\
    \url{{kalofolias,mboley,jilles}@mpi-inf.mpg.de}
  }
}
\fi
\begin{document}

\setlength{\pdfpagewidth}{8.5in}
\setlength{\pdfpageheight}{11in}

\title{\ourmaintitle}

\date{}

\maketitle
   
\begin{abstract}
Subgroup discovery is a local pattern mining technique to find interpretable descriptions of sub-populations that stand out on a given target variable. That is, these sub-populations are exceptional with regard to the global distribution. In this paper we argue 
that in many applications, such as scientific discovery, subgroups are only useful if they are additionally representative of the global distribution with regard to a control variable. That is, 
when the distribution of this control variable is the same, or almost the same, as over the whole data.

We formalise this objective function and give an efficient algorithm to compute its tight optimistic estimator for the case of a numeric target and a binary control variable. This enables us to use the branch-and-bound framework to efficiently discover the top-$k$ subgroups that are both exceptional as well as representative. Experimental evaluation on a wide range of datasets shows that with this algorithm we discover meaningful representative patterns and are up to orders of magnitude faster in terms of node evaluations as well as time.

\end{abstract}

\ifkeywords
\fi

\section{Introduction}
\label{sec:intro}
Pattern mining in general, and subgroup discovery in particular, are powerful exploratory data mining techniques that can reveal important local structure that can easily be missed, or not explicitly
represented, by global models~\cite{atzmueller_subgroup_2015}. More precisely, subgroup discovery aims to find interpretable selectors of local data regions by optimising a trade-off between exceptionality, i.e., the degree to which the distribution of a designated target variable varies locally from its global distribution, and generality, i.e., the fraction of the data space covered by the selector.

A problem with this traditional approach is its simplistic notion of generality: if a subpopulation is relatively sizeable it is considered general, even though it might show arbitrary statistical obscurities. 
This lack of representativeness is a key problem in many important scenarios. 

In scientific discovery and theory development, we often seek to 
identify local factors that influence some variable, but want to control 
for the influence of other potential explanations. 
For instance, in materials science we may want to discover
structural patterns that characterise the HOMO-LUMO energy gap in gold nanoclusters~\cite{goldsmith:17:uncovering}, 
independent of the parity of their atom count, which is already known to have a strong influence.
As another example, in political science we are often interested in discovering 
demographics with a high affinity to a certain political party. However, findings should not rediscover known geographic influences (See Fig.~\ref{fig:example_intro}). 

Besides science, there are other of examples where traditional subgroup discovery fails. In policy development and other fairness-aware applications there 
are often ethical and legal requirements that demand the distribution of policy recipients to match the underlying population w.r.t.\ to some sensitive
variable. For instance, while students with a high chance
of obtaining a degree are reasonable candidates for defining
the application criteria of a scholarship, we might still want to ensure
that the eligible population is gender-balanced.

\newsavebox{\boxtable}
\newsavebox{\boxsg}
\newsavebox{\boxcdf}
\newsavebox{\boxreg}
\definecolor{creedobox}{HTML}{E5EDED}
\newlength{\plotheight}
\plotheight=2.5cm
\newcommand{\drawcdf}[1][\linewidth]{%
  \begin{tikzpicture}%
    \begin{axis}[
      eda axis,
      width=#1,
      height=\plotheight,
      probability axis y,
      probability axis x,
      xlabel={\strut DIE LINKE\%},
      xtick distance=10,
      x tick label style={font=\scriptsize},
      ]
      \addplot[subgroup line] table[x=y,y=cdfQ]{\tblpatcdf};
      \addplot[population line] table[x=y,y=cdfP]{\tblpatcdf};
    \end{axis}
  \end{tikzpicture}%
}
\newcommand{\drawreg}[1][\linewidth]{%
  \begin{tikzpicture}%
    \begin{axis}[
      width=#1,
      height=\plotheight,
      eda ybar,
      y tick label style = {normal shift={y}{-3pt}}, 
      xmin=-.5,
      xmax = 1.5,
      major grid style={black, dotted},
      eda axis,clip = false,
      probability axis y,
      xtick=data,
      ymin=0,
      xlabel={\strut Region},
      bar width=6,
      legend style={
        at={(0.7,0.3)},
        legend columns=1,
        rounded corners=2pt,
        anchor=north west},
      xticklabels from table={\tblpatprb}{c},
      bar width=4pt,
      ]
      \addplot[population line,fill] table[x expr=\coordindex,y=pP]{\tblpatprb};
      \addplot[subgroup line,fill] table[x expr=\coordindex,y=pQ]{\tblpatprb};
    \end{axis}
  \end{tikzpicture}%
}
\newcommand{\drawtable}[1][\linewidth]{%
  \scriptsize%
  \pgfplotstabletypeset[scoretable]\tblpatstatstrn%
}
\newcommand{\renderSubgroup}[2]{
  \def\sgtag{#1}
  \def\sgname{#2}
  \pgfplotstableread[header=true]{./csv/pattern-\sgtag-cdf.csv}\tblpatcdf
  \pgfplotstableread[header=true]{./csv/pattern-\sgtag-prb.csv}\tblpatprb
  \pgfplotstablegetelem{0}{y}\of{\tblpatcdf}%
  \pgfmathsetmacro{\xminprb}{\pgfplotsretval}
  \pgfplotstablegetrowsof{\tblpatcdf}
  \pgfmathtruncatemacro{\lastrowprb}{\pgfplotsretval-1} 
  \pgfplotstablegetelem{\lastrowprb}{y}\of{\tblpatcdf}%
  \pgfmathsetmacro{\xmaxprb}{\pgfplotsretval} 
  \pgfplotstableread[header=true]{./csv/pattern-\sgtag-stats.csv}\tblpatstats
  \pgfplotstabletranspose[
  header=false,
  colnames from=key,
  input colnames to=key
  ]\tblpatstatstrn\tblpatstats
  {
    \colorbox{creedobox}{
      \begin{minipage}{.92\linewidth}
        \tikzset{
          subgroup line/.style={red},
          population line/.style={blue},
        }
        \pgfplotsset{
          probability axis y/.style={
            try min ticks=3,
            ymax=1,
            ytick={0,0.5,1},
            yticklabels={$0$,,{$1$}},
          },
          probability axis x/.style={
            xmin=0,
            xmax=\xmaxprb
          },
          eda axis/.style={
            eda line,
            ticklabel shift = 0,
            scaled ticks = false,
            x label style = {},
            y label style = {},
            xtick style = {font=\scriptsize},
            ytick style = {font=\scriptsize},
            x tick label style = {yshift=-3pt},
            y tick label style = {xshift=-3pt},
            xtick style = {normal shift={x}{-8pt}},
            ytick style = {normal shift={y}{-8pt}},
            axis lines*=left,
            x label style={normal shift={x}{-8pt}},
            y label style={normal shift={y}{-8pt}},
            x axis line style = {thick,normal shift={x}{-8pt}},
            y axis line style = {thick,normal shift={y}{-8pt}},
            enlargelimits = false,
          },
        }
        \pgfplotstableset{
          scoretable/.style={
            column type=,
            display columns/0/.style={string type},
            every head row/.style={output empty row},
            begin table={\begin{tabularx}{\linewidth}{@{}Xr@{=}l@{}}},
              end table={\end{tabularx}},
            create on use/desc/.style={
              create col/set list={coverage:,tendency:,represent.:},
            },
            columns/desc/.style={
              string type,
            },
            columns/key/.style={
              column name={Score},
              string type,
              preproc cell content/.append style={/pgfplots/table/@cell content/.add={$}{$}},
            },
            columns/value/.style={
              fixed,fixed zerofill,precision=3,
            },
            columns={desc,key,value},
          }
        }
        {\fboxsep=0pt
          \begin{tabu}{@{\hspace{.0\linewidth}}m{.54\linewidth}@{\hspace{.03\linewidth}}m{.44\linewidth}@{}}
            \smaller Subgroup & \smaller Scores\\[-.25ex]
            \scriptsize\texttt{\sgname} &
            \makebox[\linewidth]{%
              \drawtable%
            }
            \\[.6ex]
            \makebox[\linewidth]{
                \hspace{-1.5em}\drawcdf[1.3\linewidth]%
              }%
              &
            \makebox[\linewidth]{
                \hspace{-1.1em}\drawreg[1.5\linewidth]%
              }%
            \\[-1ex]
          \end{tabu}
        }
      \end{minipage}
    }
  }
}
\tikzexternaldisable
\begin{figure*}
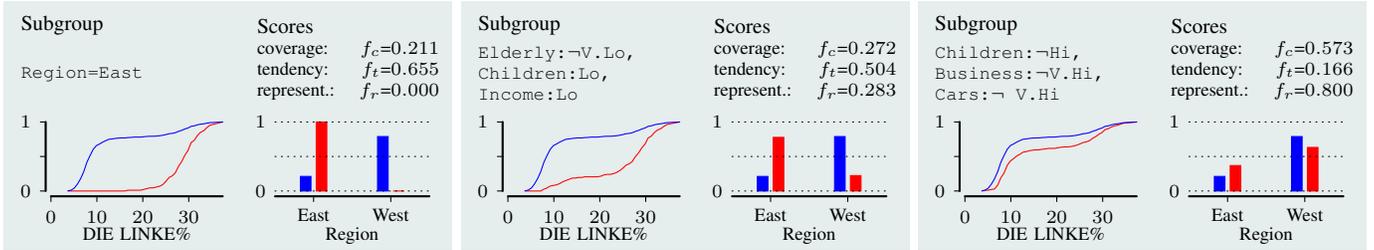

  \begin{minipage}{0.33\linewidth}
    \renderSubgroup{region}{Region=East}
  \end{minipage}
  \begin{minipage}{0.33\linewidth}
    \renderSubgroup{missing}{%
      Elderly:$\neg$V.Lo, Children:Lo, Income:Lo}
 \end{minipage}
  \begin{minipage}{0.33\linewidth}
    \renderSubgroup{controlled}{Children:$\neg$Hi, Business:$\neg$V.Hi, Cars:$\neg$ V.Hi}
  \end{minipage}
  \caption{Subgroups of German voting districts in 2009 elections with high percentage of the left-wing party ``DIE LINKE''. Blocks show the cdf of percentage in the subgroup (red) and global population (blue), along with the the distribution of district locations. Traditional subgroup discovery (left) recovers the main trend: eastern districts support ``DIE LINKE''. Removing the \emph{region} attribute (middle) results in a similar subgroup. Only when explicitly controlling for geography (right) do we discover subgroups that stand out with regard to voting behaviour, while at the same time being representative for the whole country.
}\label{fig:example_intro}
\end{figure*}
\tikzexternalenable

All of the above settings share the requirement of subgroups 
to not only be relatively sizeable, but also statistically \emph{representative} 
w.r.t.\ some control variable. Specifically, this variable should have a similar distribution between the subgroup and the global population, exhibiting what is called \temph{statistical parity}~\cite{zemel_learning_2013}.
In contrast, simply removing the control variable, to avoid it influencing the result, is infeasible, since it can usually be approximately recovered by the remaining variables
(known as \emph{red-lining effect}~\cite{calders_three_2010}). See again Fig.~\ref{fig:example_intro}.

While there are several techniques to enforce representativeness of binary global 
classifiers, it is unclear how those can (i) be generalised to settings that go beyond a binary prediction task, and (ii) be integrated effectively into branch-and-bound,
the standard framework for optimal subgroup discovery. This framework requires an efficiently, i.e., near linear time, computable optimistic estimator for the desired objective function.

Therefore, in this work we propose a general representativeness term that can be incorporated into subgroup discovery objective functions, which is based on the statistical distance between the local and the global distribution of the control variable. Moreover, we show how the resulting representative subgroup discovery problem can be solved efficiently for the case of a binary control and a numeric target variable.
In particular, we propose \acs{RAWR}, an algorithm to compute the tight optimistic estimator for the representativeness-aware objective function, in $O(n \log n)$ time.
Experiments show that, when employing this algorithm in the branch-and-bound framework, we can prune orders of magnitude of candidates in comparison to the state of the art, which, besides reducing memory consumption, leads to orders of magnitude gain in runtime; therewith, \acs{RAWR} makes it possible to mine representative subgroups in otherwise computationally infeasible settings.

\section{Preliminaries}\label{sec:prelim}
In this section we recall the necessary basics of subgroup discovery and \tBNB search.

\subsection{Subgroup discovery}\label{sec:subgroup-values}
The goal in subgroup discovery is to identify useful subpopulations of a given \tdef{global population} $\sPop$, which can be viewed as a set of $\snPop[*]$ entities $\sPop=\{\sitem_1,\ldots,\sitem_{\snPop[*]}\}$. 
These sub-populations are identified by Boolean functions, $\sfmap\sLMem\sPop\fBools$, the \tdef{subgroup selectors}, each of which defines a subpopulation $\sSub=\sfext(\sLMem)$ through the extension function $\sfext:\sLMem\mapsto \{\sitem\in\sPop : \sLMem(\sitem)=\top\}$; note that we will often use $\sLMem$ and $\sSub$ interchangeably.
The set of all available subgroup selectors, the \temph{selection language} $\sLang$, most commonly comprises conjunctions formulated over a set of basic descriptive conditions, e.g., \tpred{[age>18]} or \tpred{[sex=\textnormal`Male\textnormal']}.
In this paper, however, it suffices to consider an abstract selection language.

Additionally, we assume a continuous \tdef{target variable} $\sfmap \svtar\sPop\fReals$ and a discrete \tdef{control variable} $\sfmap\svctr\sPop{\{1,\dots,\snCls\}}$.
The usefulness of a subgroup can then be encoded by a real-valued \tdef{objective function} $\sfmap \sffit {2^\sPop} \fReals$.
An exemplary such function, for numeric target variables, is the \temph{impact function} 
\begin{align}
  \label{eq:objective-other}
  \sCovCent(\sSub)\eqdef
  \sCov(\sSub) \sCent(\sSub) =
  \frac{\sfcard\sSub}{\sfcard\sPop}
  \frac{\favg[\sSub]{\svtar}-\favg[\sPop]{\svtar}}{\max_{\sitem \in \sPop} \svtar(\sitem) - \favg[\sPop]{\svtar}} \enspace,
\end{align}
where $\favg[\sPop]\svtar\eqdef\mean\{\svtar(\sitem):\sitem\in\sPop\}$ is the mean of the target values in the population, and $\favg[\sSub]\svtar$ is the mean of those in $\sSub$.

A subgroup $\sSub$ with a high $\sCovCent$ value is exceptional, as the \tdef{central tendency factor}
\begin{align}
   \label{eq:central-tendency-definition}
   \sCent(\sSub)\eqdef\frac{\favg[\sSub]{\svtar}-\favg[\sPop]{\svtar}}{\max_{\sitem \in \sPop} \svtar(\sitem) - \favg[\sPop]{\svtar}}
\end{align}
ensures a high mean deviation of $\svtar$ within the subgroup. At the same time, $\sSub$ exhibits a basic notion of generality, provided by the \tdef{coverage factor} $\sCov(\sSub)\eqdef |\sSub|/|\sPop|$.

In Sec.~\ref{sec:theory} we will augment this objective function to also represent a statistical notion of generality of $\sSub$ w.r.t.\ the control variable.

\subsection{Branch and Bound with Optimistic Estimators}

The standard algorithm for finding a set of k optimal subgroup selectors is \acl{BNB} (here we give a basic overview and refer to Boley et al.~\cite{boley_identifying_2017} for more details).
This algorithm employs a refinement operator $\sLRef:\sLang \longrightarrow 2^\sLang$, as well as an \temph{optimistic estimator} $\sfoest$ of the objective function $\sffit$,
\begin{align}
	\label{eq:optimistic-estimator-maximize}
	\sfoest(\sSub) \geq 
	\max_{\substack{\sSel\subseteq \sSub, \sSel\neq \emptyset}}\sffit(\sSel), &&\forall\sSub\subseteq\sPop\enspace.
\end{align}

The algorithm maintains a priority queue of candidate subgroup selectors $\sLMem\in\sLang$, initialised to contain only a root selector, describing the entire population $\sPop$. While keeping track of the subgroup $\sSub^*$ with the the best value discovered so far, the algorithm iterates by picking from the queue that selector (resp.\ subgroup) $\sSub=\sfext(\sLMem)$ with the highest optimistic estimator value $\sfoest(\sSub)$; this favours subgroups with the greatest potential for improvement, resulting in a best-first-search scheme. If the optimistic estimator of $\sSub$ ensures that none of its subgroups surpass the current best, i.e., $\sfoest(\sSub) < \sffit(\sSub^*)$, all its refinements $\sLRef(\sLMem)$ can be safely pruned; otherwise, these refinements are placed in the queue. This procedure iterates until the queue empties, which guarantees to find the best, or by an easy extension, the best $k$ subgroups.

Obviously, as the bound of the optimistic estimator gets tighter, the pruning potential increases, to become maximal when Eq.~\eqref{eq:optimistic-estimator-maximize} holds with equality. Then we refer to $\sfoest$ as a \tdef{tight optimistic estimator}~\cite{grosskreutz_tight_2008} of the objective function $\sffit$.

Optionally, and to achieve even better pruning, the \tBNB algorithm may use the relaxed comparison $\sappr\hat\sffit(\sSub) > \sffit(\sSub^*) $, for an approximation factor $\sappr \in (0,1]$, where a value of $\sappr=1$ yields the best subgroup. Lower $\sappr$ generally yield better pruning, while guaranteeing that the discovered subgroup has a value no less than $\sappr$ times that of the best subgroup.

The impact function $\sCovCent$ allows an efficient implementation of its tight optimistic estimator, (since $\max_{\sSub\subset \sPop} \sCovCent(\sSub)=\sCovCent(\sSub^*)$, with $\sSub^*=\{\sitem: \svtar(e) \geq \favg[\sPop]{\svtar}\}$
), computable in linear time~\cite{boley_identifying_2017,lemmerich2016fast}. We refer to this implementation as \tBRIG, to remind that $\sCovCent$ is oblivious to the control variable.

In the next section we develop \tRAWR, an efficient algorithm to compute the tight optimistic estimator for the controlled impact function, and also show that any optimistic estimator of the impact function (and thus also \tBRIG) can be used as a non-tight stand-in for the augmented one.

\section{Representative Subgroup Discovery}
\label{sec:theory}

In order to describe the theoretical contributions of this paper, let us fix the following notation. We consider subpopulations $\sSub\subseteq\sPop$, whose items we assume to be ordered in decreasing target value. Hence, $\svtar_i$ is the item of $\sSub$ with the $i$-th greatest target value, which has a control class of $\svctr_i$. Out of those elements of $\sSub$ with class $\siCls$, we denote $\sfvtar\siSub$ the one with the $\siSub$-th greatest target value, and by $\fnSet[\siCls]\sSub\eqdef|\{\sitem\in\sSub: \svctr(\sitem)=\siCls\}|$ their count, which we also refer to as the $\siCls$-th \temph{class count}. Similarly, we define the \temph{class probability vector} $\fpSet[]\sSub$ with elements the \temph{class probabilities} $\fpSet[\siCls]\sSub\eqdef \fnSet[\siCls]\sSub/|\sSub|$, for each class $\siCls$.

\subsection{The controlled impact function}\label{sec:representativeness}
We now augment the standard objective function of Eq~\eqref{eq:objective-other} to also account for a broader notion of generality than coverage: the statistical generality of the subgroup w.r.t.\ the control variable. Specifically, we add a \tdef{representativeness factor} $\sRepr(\sSub)$, quantifying the similarity of the control distribution between $\sSub$ and $\sPop$. This forms the \temph{controlled impact function}
\begin{align}
  \label{eq:objective-ours}
  \sffit(\sSub)\eqdef
  \sCovCent(\sSub)^{1-\sweight}
  \sRepr(\sSub)^\sweight\enspace,
\end{align}
where the $\sweight\in[0,1)$ parameter tunes the trade-off between representativeness and the typical properties quantified by $\sCovCent$.

Viewed probabilistically, our goal is to select subpopulations independently of the control variable. This is equivalent to requiring, for a random entity $\sritem\in\sPop$ from the population, that
\begin{align}
  \sProb\big(\svctr(\sritem) | \sLMem(\sritem)=\top\big)
  = \sProb\big(\svctr(\sritem)\big) \iff
  \sfdist(\spSub,\spPop)=0\;,
\end{align}
where $\sfdist$ is some distance measure between distributions with $\spSub\eqdef\fpSet[]\sSub$ and $\spPop\eqdef\fpSet[]\sPop$. In this work, we further fix $\sfdist$ to be the \temph{total variation distance} $\sfdist(\spSub,\spPop) = \frac 1 2\sum_\siCls |\spSub[\siCls]-\spPop[\siCls]|$, equal to the maximal difference between probabilities of any set of control classes. This measure is at once intuitively interpretable and simple enough to allow efficient calculations.

Similar to the coverage and the central tendency factors, we design this new factor to assume values in the interval $\sRepr\in[0,1]$, with more representative subgroups scoring higher:
\begin{align}
  \label{eq:representativeness}
  \sRepr(\sSub) \eqdef 1-\frac{\sfdist(\spPop,\spSub)-\sfdist[max]}{\sfdist[max]}, && 
                                                                                      \sfdist[max] \eqdef \displaystyle\max_{\sSel\subseteq\sPop}\sfdist(\spPop,\spSel)\;.
\end{align}
We note that an important consequence of these bounds is that any optimistic estimator for the impact function is also a valid, albeit non-tight, optimistic estimator for the controlled impact function. This corresponds to fixing the added factor to its maximal value of $\sRepr=1$.

Having introduced all constituents of the controlled impact function, we now proceed with the computation of its tight optimistic estimator. We first introduce a transform of the domain of the original optimisation problem from exponential to polynomial size in Sec.~\ref{sec:ccs}. We then employ this transform in Sec.~\ref{sec:special-case} to derive an efficient algorithm that computes this tight optimistic estimator in $O(n\log n)$ time, for the special case of a population with balanced binary classes.

\subsection{Searching in the \tCCS}\label{sec:ccs}

In this section we describe a transformation which aggregates the exponentially many subsets of $\sSub$ in the original optimisation problem of Eq.~\eqref{eq:optimistic-estimator-maximize} into polynomially many sets of subsets. Additionally, the maximum $\sffit$ value attained by any subset within each of these sets can be efficiently computed. From now on, we call any subset $\sSel\subseteq\sSub$ a \temph{refinement} of $\sSub$.

For any given subgroup $\sSub$, we consider the space of all possible class count vectors $\sPInd\eqdef\big(\fnSet[1]\sSub,\ifmce{\fnSet[2]\sSub}{\ldots,\fnSet[\snCls]\sSub}\big)$ that any refinement $\sSel\subseteq\sSub$ might assume,
  \begin{align}
    \label{eq:class-counting-space}
    \sPIndSet(\sSub) &
                       \eqdef
                       \ifmce{
                       \{0,\ldots,\fnSet[1]{\sSub}\}\times\{0,\ldots,\fnSet[2]{\sSub}\}}{
                       }{\bigtimes_{\siCls=1}^\snCls\{0,\ldots,\fnSet\sSub\}}\enspace.
  \end{align}
  This space, which we refer to as the \temph{\acf{CCS}}, is a subset of the \ifmce{square lattice $\mathbb{Z}^2$}{lattice $\mathbb{Z}^\snCls$}, and partitions the original space $2^\sSub$ into $|\sPIndSet(\sSub)|=\ifmce{(\fnSet[1]{\sSub}+1)\cdot(\fnSet[2]{\sSub}+1)}{\prod_{\siCls=1}^\snCls(\fnSet\sSub+1)}$ partitions. Each of these partitions, called the \tdef{equi-count refinement sets} $\sSelPar(\sSub)$, consists of these refinements of $\sSub$ with \ifmce{$\sPInd[1]$ and $\sPInd[2]$ items of control class $1$ and $2$, respectively}{$\sPInd[\siCls]$ items of control class $\siCls$, for each class $\siCls=1,\ldots,\snCls$},
\begin{align}
  \label{eq:equi-count-partition}
  \sSelPar(\sSub) \eqdef \{\sSel\subseteq\sSub: \ifmce{\fnSet[1]\sSel=\sPInd[1],\;\fnSet[2]\sSel=\sPInd[2]}
  {\fnSet[\siCls]\sSel=\sPInd[\siCls],\quad \siCls=1,\ldots,\snCls}
  \}\;.
\end{align}
For an example of a \tCCS \ifmc{with $\snCls=2$ classes} see Fig.~\ref{fig:toy}.

\begin{figure}
  \begin{subfigure}{\columnwidth}
    \centering
    \resizebox{\linewidth}{!}{
\begin{tikzpicture}
  \begin{axis}[
    clip mode=individual,
    small ticks,
    width=\linewidth,
    height=0.36\linewidth,
    ylabel near ticks,
    every axis x label/.style=
    {at={(ticklabel cs: 0.5,0)}, anchor=north},
    legend style={legend columns=1,at={(.95,0.625)},anchor=east,font=\smaller},
    ymin=0.8,
    ymax=2.2,
    xmin=-1,
    xmax=2,
    ytick={1,2},
    xlabel=Target (\svtar),
    ylabel=Control (\svctr)
    ]
    \addplot[only marks,mark=*,mark size=1pt,blue] coordinates {
      (-0.877916,1)
      (-0.793658,1)
      (-0.516421,2)
      (-0.383239,1)
      (-0.190260,1)
      (0.176670,2)
      (0.592525,2)
      (0.774632,2)
      (0.894632,2)
      (1.084485,1)
      (1.217863,1)
      (1.435129,1)
      (1.673677,1)
    };
    \addplot[only marks,mark=o,mark size=2pt,green!50!black] coordinates {
      (-0.877916,1)
      (-0.516421,2)
      (-0.383239,1)
      (-0.190260,1)
      (0.176670,2)
      (0.774632,2)
      (0.894632,2)
      (1.435129,1)
      (1.673677,1)
    };
    \addplot[only marks,mark=square,mark size=3.5pt,red] coordinates {
      (-0.877916,1)
      (-0.516421,2)
      (0.176670,2)
      (0.774632,2)
      (1.435129,1)
    };
    \legend{\sPop,\sSub,\sSel};
  \end{axis}

\end{tikzpicture}

    }
    \caption{
      Toy population \sPop, consisting of $\fnSet[1]\sPop=8$ items of control class $1$ and $\fnSet[2]\sPop=5$ of class $2$, a subgroup $\sSub\subseteq\sPop$, with $\fnSet[1]{\sSub}=5$ items of class $1$ and $\fnSet[2]\sSub=4$ of class $2$, and a refinement $\sSel\subset\sSub$.\label{fig:toy-dataset}}
  \end{subfigure}
  \begin{subfigure}{\columnwidth}
    \centering
    \resizebox{\linewidth}{!}{
\begin{tikzpicture}
  \def\xmax{8}
  \def\ymax{5}
  \begin{axis}[
    small ticks,
    clip mode=individual,
    width=\linewidth,
    height=0.625\linewidth,
    axis lines=middle,
    ylabel near ticks,
    xlabel near ticks,
    ymin=0,
    ymax=\ymax+1,
    xmin=0,
    xmax=\xmax+1,
    xtick={0,...,\xmax},
    ytick={0,...,\ymax},
    xlabel={$\fnSet[1]\sSel$},
    ylabel={$\fnSet[2]\sSel$},
    legend style={at={(1,0.1)},nodes={scale=1, transform shape},anchor=south east,font=\smaller}
    ]
    
    \foreach \x in {0,...,5}{
      \foreach \y in {0,...,4} {
        \edef\temp{
          \noexpand\node[draw,circle,point] at (axis cs:\x,\y) {};
        }
        \temp
      }
    }
    \addlegendimage{only marks,mark=o,mark options={scale=1}};
    \addlegendentry{points of $\sPIndSet(\sSub)$}
    
    \addplot[repr,only marks,mark=*] coordinates {(\xmax,\ymax)};
    \addlegendentry{class count of $\sPop$};
    
    \addplot[repr] coordinates {(\xmax,\ymax) (0,0)};
    \addlegendentry{Max $\sRepr$ ray}
    
    \node[draw=none,circle,point] at (axis cs:2,3) (ref) {};
    \path[draw,red,>=stealth] (axis cs: 2.5,4.5) node[above] {$\sSelPar[(2,3)]$} edge[out=260,in=50,->] (ref);
    \node[draw=none,circle,point] at (axis cs:8,5) (ref2) {};
    \path[draw,blue,>=stealth] (axis cs: 7.5,3.5) node[scale=0.8] {$(\fnSet[1]\sPop,\fnSet[2]\sPop)$} edge[out=70,in=270,->] (ref2);
  \end{axis}
\end{tikzpicture}

    }
    \caption{The \acs{CCS} of $\sSub$, denoted $\sPIndSet(\sSub)$, and the maximum $\sReprCCS{}$ ray. Each refinement $\sSel\subseteq\sSub$ with class counts $\sPInd=(\sPInd[1],\sPInd[2])$, where $\sPInd[1]=\fnSet[1]\sSel$ and $\sPInd[2]=\fnSet[2]\sSel$, is contained in the equi-class refinement set $\sSelPar$, which corresponds to the point $\sPInd$ in $\sPIndSet(\sSub) = \{0,\ldots,5\}\times\{0,\ldots,4\}$. Points closer to the \emph{max $\sRepr$ ray} have a class count probability (ratio) closer to that of $\sPop$ and thus a higher $\sRepr$ score.
    }\label{fig:toy-ccs}
  \end{subfigure}
  \caption{The \acl{CCS} (bottom) for a toy population \ifmc{with $\snCls=2$ control classes }(top). The refinement $\sSel$ is contained in $\sSelPar[(2,3)]$, corresponding to the annotated point.
  }\label{fig:toy}
\end{figure}
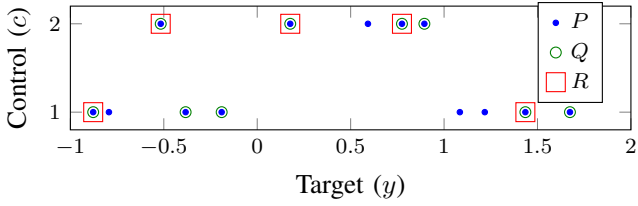
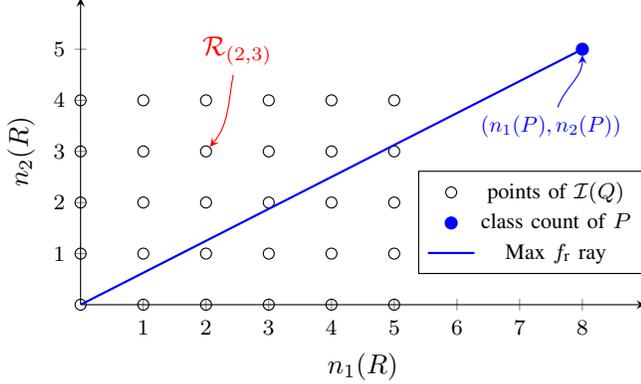

The computation of the tight optimistic estimator $\sfoest(\sSub) \eqdef \max_{\sSel \in 2^\sSub}\sffit(\sSel) $ of Eq~\eqref{eq:optimistic-estimator-maximize} can now be expressed as
\begin{align}
  \label{eq:objective-optimisation-task}
  \sfoest(\sSub) \eqdef
  \max_{\sPInd \in \sPIndSet(\sSub)}
  \max_{\sSel \in \sSelPar(\sSub)}
  \sffit(\sSel)
  =
  \max_{\sPInd \in \sPIndSet(\sSub)}\sffitCCS\sPInd\;,
\end{align}
where $\sffitCCS\sPInd$ refers to the maximal value attained over all refinements in the equi-count refinement set $\sSelPar$
\begin{align}
  \label{eq:objective-over-ccs}
  \sffitCCS\sPInd \eqdef
  \begin{cases}
    \displaystyle \max_{\sSel\in\sSelPar(\sSub)}\sffit(\sSel)& \sPInd\in\sPIndSet(\sSub)\setminus \{\fVec0\}\\
    -\infty& \sPInd = \fVec0
  \end{cases}\;.
\end{align}
Similarly, the maxima of the impact function, central tendency and representativeness values over all refinements within $\sSelPar$ are denoted $\sCovCentCCS\sPInd$, $\sCentCCS\sPInd$ and $\sReprCCS\sPInd$, respectively.

In the next proposition we derive a closed form for the optimiser of $\sffit(\sSub)$ within an equi-count refinement set $\sSelPar$, which can then be used to compute $\sCovCentCCS\sPInd$ and thus $\sffitCCS\sPInd$.

\begin{proposition}
  \label{thm:efficient-naive-opt}
  The optimal value $\sCovCentCCS\sPInd$ is attained by the set
  \begin{align}
    \label{eq:optimal-partition-subset}
    \sSelParOpt\eqdef
    \ifmce{
    \left\{\sfvtar[1]1,\ldots,\sfvtar[1]{\sPInd[1]}\right\}
    \cup
    \left\{\sfvtar[2]1,\ldots,\sfvtar[2]{\sPInd[2]}\right\}
    }
    {\bigcup_{\siCls=1}^\snCls\left\{\sfvtar1,\ldots,\sfvtar{\sPInd[\siCls]}\right\}}
    \enspace,
  \end{align}
   which contains the $\sPInd[\siCls]$ items with the greatest target value among those with control class $\siCls$, for \ifmce{both classes $\siCls=1,2$}{all classes $\siCls=1,\ldots,\snCls$}.
  
  \proof
  All sets $\sSel \in \sSelPar(\sSub)$ have a constant coverage $|\sSel|=\ifmce{\sPInd[1]+\sPInd[2]}{\sum_{\siCls=1}^\snCls\sPInd[\siCls]}$, so that maximising the objective value is equivalent to maximising the central tendency factor $\sCentCCS{}$. We will show that $\sSelParOpt$ attains the greatest $\sCent$ value over $\sSelPar(\sSub)$ by contradiction.

  Assume there is a refinement $\sSel'\in\sSelPar$ with $\sSel'\neq \sSelParOpt$ and $\sCent(\sSel')>\sCent(\sSelParOpt)$. Since $\sSelParOpt$ contains the items with maximum $\svtar$ value for each class, there is at least one sequence of refinements $\big(\sSel^{(0)},\ldots,\sSel^{(T)}\big)$, starting with $\sSel^{(0)}\eqdef\sSelParOpt$ and ending at $\sSel^{(T)}\eqdef\sSel'$, so that at each index $\tau$ we exchange a single element between $\sSel^\tau$ with another in $\sSub\setminus \sSel^\tau$ of the same class, but a smaller target value. Formally, $\sSel^{(\tau)} = \big(\sSel^{(\tau-1)}\setminus\{\sitem\}\big)\cup\{\sitem'\}$, such that $\svctr(\sitem')=\svctr(\sitem)$ and $\svtar(\sitem')<\svtar(\sitem)$. This implies that, for each $\tau=2,\ldots,T$ we get
  \begin{align}
    \sum_{\sitem\in\sSel^{(\tau)}}\svtar(\sitem) - \sum_{\sitem\in\sSel^{(\tau-1)}}\svtar(\sitem)=
    \svtar(\sitem')-\svtar(\sitem) < 0\;.
  \end{align}
Dividing these sums with \ifmce{$\sPInd[1]+\sPInd[2]$}{$\sum_{\siCls=1}^\snCls\sPInd[\siCls]$}, turns them into means, and since $\sCent$ is increasing w.r.t.\ the target value mean, we have $\sCent(\sSel^{(\tau)})<\sCent(\sSel^{(\tau-1)})$. By transitivity, it is $\sCent(\sSel')<\sCent(\sSelParOpt)$, contradicting the optimality of $\sCent(\sSel')$.\qed
\end{proposition}

As a result, we can express the target value mean of the optimiser $\sSelParOpt$ as 
$
   \mean(\sSelParOpt)=
   \sum_{\siCls=1}^{\ifmce2\snCls}
   \sum_{\siSub=1}^{\sPInd[\siCls]}\sfvtar\siSub/\|\sPInd\|_1
$,
where $\|\sPInd\|_1=\ifmce{\sPInd[1]+\sPInd[2]}{\sum_{\siCls=1}^\snCls\sPInd[\siCls]}$ is the cardinality of each refinement in $\sSelPar$. Now the impact function $\sCovCent$ of Eq.~\eqref{eq:objective-other} can be transformed onto the \tCCS as
\begin{align}
  \label{eq:objective-covcent-cspace}
  \sCovCentCCS\sPInd\eqdef& \sconstt \sum_{\siCls=1}^{\ifmce2\snCls}\sum_{\siSub=1}^{\sPInd[\siCls]}\sfvtar\siSub -\sconstc\|\sPInd\|_1\;,
\end{align}
where
\begin{align}
 \sconstt=\frac1{\sDenom{}}>0\,,\;
 \sconstc=\frac{\favg[\sPop]\svtar}{\sDenom{}}\,,
 \text{ and }\;
 \sDenom{}=|\sPop|\Big(\max_{\sitem\in\sPop}\svtar(\sitem)-\favg[\sPop]\svtar\Big)\;.
\end{align}

Since the representativeness factor $\sRepr(\sSub)$ depends only on the class counts of $\sSub$, it remains constant over $\sSelPar$ and does not affect the maximiser. Therefore, the transformed controlled impact function can be written as
\begin{align}
  \label{eq:objective-special-over-ccs}
  \sffitCCS\sPInd \eqdef
{\sCovCentCCS\sPInd}^{1-\sweight}\cdot{\sReprCCS\sPInd}^\sweight\; && \sweight\in[0,1)\enspace.
\end{align}

Notice that the value $\sCovCentCCS\sPInd$ can be computed in constant time for any point $\sPInd\in \sPIndSet(\sSub)$, after a pre-processing step of linear time. Indeed, assuming the items of a candidate subgroup are in decreasing order of target values, we can achieve this by first passing through the values and creating \ifmce{$2$}{$\snCls$} cumulative sums of target values, one for each class; after this step, the value $\sCentCCS\sPInd$ can be easily retrieved as the sum of indices $\sPInd[\siCls]$ within the cumulative sum for each class $\siCls$, appropriately scaled. The controlled impact function $\sffitCCS\sPInd$ can be computed with trivial extra work to compute $\sReprCCS\sPInd$.

\ifmce{Note that the entire analysis so far can be easily extended to populations with arbitrary discrete controls. This allows the computation of a tight bound for $\sffit$ on such populations, by iterating in  $O(n^\snCls)$ time over the entire \acs{CCS}, now $\sPIndSet(\sSub)\subset \mathbb{Z}^\snCls$.}{}

Therefore, this transform can be used in a straightforward way to derive an algorithm to compute the tight optimistic estimator in $O(n^{\ifmce2\snCls})$ time. However, a practical algorithm can benefit from further improvement, achieved in the next section for a special case of a population.

\subsection{A linearithmic algorithm for balanced binary controls}\label{sec:special-case}

We now present a linearithmic algorithm able to compute the tight optimistic estimator of the controlled impact function of Eq.~\eqref{eq:objective-ours} for the case of a population $\sPop$ with balanced binary control classes, i.e., $\sfmap\svctr\sPop{\{1,2\}}$ with $\fnSet[1]\sPop=\fnSet[2]\sPop$.

The rest of the analysis can be summarised in two key steps. First, we show that there is a sub-region of $\sPIndSet(\sSub)$ where $\sffitCCS\sPInd$ attains its maximum and then we present an efficient algorithm to search within this sub-region.

For this purpose we study the two factors, $\sCovCentCCS{}$ and $\sReprCCS{}$ within the \tCCS.
\begin{figure}
  \resizebox{\linewidth}{!}{
    \figWidth=1.1\linewidth
    \figHeight=1.1\linewidth
\begin{tikzpicture}
  \begin{axis}[%
    small ticks,
    ccs plot 2,
    width=\figWidth,
    height=\figHeight,
    legend style={at={(0.965,0.025)},legend cell align=left,align=left,draw=black,anchor=south east,font=\smaller},
    ]
    \foreach \x in {0,...,10}{
      \foreach \y in {0,...,10}{
        \edef\temp{
          \noexpand\node[draw,circle,point] at (axis cs:\x,\y) {};
        }
        \temp
      }
    }
    \pgfplotsinvokeforeach{0,...,10}{
      \draw [concave sequence] (axis cs:0,#1) -- (axis cs:#1,#1);
      \draw [concave sequence] (axis cs:#1,0) -- (axis cs:#1,#1);
    }
    \addplot [color=blue,solid]
    table[row sep=crcr]{%
      0	0\\
      10	10\\
    };
    \addlegendentry{Max. \sReprCCS{} ray};
    
    \addplot [solid,line width=2.0pt,ctpath]
    table[row sep=crcr]{%
      0	0\\1	0\\1	1\\1	2\\1	3\\2	3\\2	4\\2	5\\3	5\\
      3	6\\3	7\\4	7\\5	7\\5	8\\6	8\\7	8\\8	8\\9	8\\
      10	8\\10	9\\10	10\\
    };
    \addlegendentry{Optimal c-t path $\sPath{}$};
    \addplot [only marks,mark=*,ctcount]
    table[row sep=crcr]{%
      4	7\\
    };
    \addlegendentry{Optimal c-t count $\sPath*$};

    \addplot [color=red,solid,only marks,mark=o,sst]
    table[row sep=crcr]{%
      4 7\\5 7\\6 7\\7 7\\
      4 6\\5 6\\6 6\\
      4 5\\5 5\\
      4 4\\
    };
    \addlegendentry{Sufficient Search Triangle $\sSST[]$};
    \addplot [sst,forget plot]
    table[row sep=crcr]{%
      4 7\\4 4\\7 7\\4 7\\
    };
    \addlegendimage{concave sequence}
    \addlegendentry{Concave sequences of $\sffitCCS{}$};
    \node[label] at (axis cs:5.5,6.5) {$\sSST$};
    \node[below=1ex,label] at (axis cs:4,4) {$(\sPath[1]*,\sPath[1]*)$};
    \node[above=2ex,label] at (axis cs:4,7) {$\sPath*=(\sPath[1]*,\sPath[2]*)$};
    \node[right=1ex,label] at (axis cs:7,7) {$(\sPath[2]*,\sPath[2]*)$};
  \end{axis}
\end{tikzpicture}

  }
  \caption{\acl*{SST} $\sSST[]$ (red circles) in the \tCCS $\sPIndSet$, and optimal point $\sPath*$ along the c-t path (crooked line), which defines the 3 vertices of $\sSST$. We seek $\sfoest=\max\sffitCCS{}$, which lies on $\sSST$, and ternary search finds it efficiently along the concave sequences of $\sffitCCS{}$ (vertical/horizontal lines).}\label{fig:ct-path}
\end{figure}
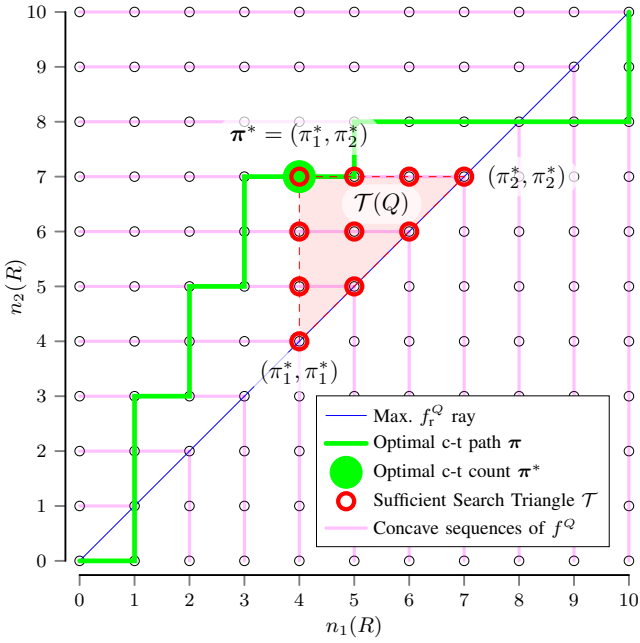
Both these factors form sequences that exhibit an appropriate notion of convexity for sequences, borrowed mutatis mutandis from Yucer~\cite{yuceer_discrete_2002}: a sequence $\sSeqA:\sSeqDom\to \fReals$, with $\sSeqDom=\{0,\ldots,\snSeq\}$ and $\snSeq\leq\infty$, is called a \tdef{convex sequence} over $\sSeqDom$, if for all $x,y\in\sSeqDom$ and each $\lambda\in(0,1)$
  \begin{align}
    \label{eq:convex-sequence}
    \lambda\sSeqA[x]+(1-\lambda)\sSeqA[y] \geq \min_{u\in\lfloor z\rfloor}\sSeqA[u]\,,&& z=\lambda x+(1-\lambda)y\;.
  \end{align}
Further, we call $\sSeqA$ a \tdef{concave sequence} if $-\sSeqA$ is convex.

We now study the $\sCovCentCCS{}$ values, as $|\sSub|=\sPInd[1]+\sPInd[2]$ increases.
 \begin{definition}[Optimal c-t path on $\sPIndSet(\sSub)$]\label{def:optimal-ct-path}
Let $\sPath\siCnt\in \sPIndSet(\sSub)$ be the maximiser of the \sCovCentCCS{} value among all points in the \tCCS with a fixed sum $\siCnt$
\begin{align}
  \label{eq:ct-path-definition}
  \sPath\siCnt \eqdef \argmax_{\sPInd\in\sPIndSet,\;\ifmce{\sPInd[1]+\sPInd[2]}{\|\sPInd\|_1}=\siCnt}\sCovCentCCS\sPInd\;, && 0\leq \siCnt\leq |\sSub|\;.
\end{align}
We refer to the optimal point sequence $\sPath{}=(\sPath 0,\ldots,\sPath{{|\sSub|}})$ as the \tdef{optimal c-t path}.
\end{definition}
The optimal c-t path exhibits useful properties, discussed in the following lemma \proofat{sec:proof-optimal-ct-path}.
\begin{lemma}
  \label{thm:optimal-ct-path}
 Let $\sSBVec[1]=(1,0)^T$ and $\sSBVec[2]=(0,1)^T$ be the standard basis vectors of $\fReals[2]$. Then (i) the $\siCnt$-th element of the optimal c-t path is the class count of the first $\siCnt$ elements of $\sPop$; formally,
  \begin{align}
    \label{eq:ct-path-solution}
    \sPath\siCnt =
    \sum_{i=1}^\siCnt \sSBVec[\svctr_i]
    &&
       0<\siCnt\leq |\sSub|
    &&
       \text{and}
    &&
       \sPath\siCnt = \fVec 0\;.
  \end{align}
Moreover, (ii) the sequence $\sCovCentCCS{}\circ\sPath{}$, with elements the $\sCovCentCCS{}$ values computed along the c-t path $\sPath{}$, is a concave sequence.
\end{lemma}

Lemma~\ref{thm:optimal-ct-path} allows for an $O(\log n)$ algorithm to find the \tdef{optimal c-t point} $\sPath*\eqdef\argmax_{\sPInd\in\sPIndSet(\sSub)}\sCovCentCCS\sPInd$, as we call the point f the \tCCS that maximises the $\sCovCentCCS{}$ value. Indeed,
\begin{align}
  \!\sCovCent(\sPath*) =
  \max_{0\leq \siCnt\leq |\sSub|}
  \max_{\substack{\sPInd\in\sPIndSet(\sSub),\,|\sPInd|=\siCnt}}\sCovCent(\sPInd)=
  \max_{0\leq \siCnt\leq |\sSub|}\sCovCent(\sPath\siCnt)\;,\!
  \end{align}
  where the last maximum runs over the $\sCovCentCCS{}$ values of the optimal path sequence. Due to the concavity of this sequence, its maximum can be computed in $O(\log n)$ time, using for instance the \temph{ternary search} algorithm.

  We now study the representativeness factor $\sRepr(\sSub)$, whose transform on the \tCCS for balanced binary controls becomes
\begin{align}
  \label{eq:tvd-ccs}
  \sReprCCS\sPInd \eqdef
  1-\left|1 - \frac{2\sPInd[1]}{\sPInd[1]+\sPInd[2]}\right|=
  1-\left|1 - \frac{2\sPInd[2]}{\sPInd[1]+\sPInd[2]}\right|\enspace.
\end{align}
We observe that the subgroups $\sSel\in\sSub$ that maximise this factor must have the same control class distribution as the population. Therefore, these subgroups must have an equal control class count $\fnSet[1]\sSel=\fnSet[2]\sSel$, and thus belong to those equi-count refinement sets $\sSelPar$, for which $\sPInd[1]=\sPInd[2]$. These, in turn, lie on the so-called \tdef{maximum $\sReprCCS{}$ ray} $\sPInd=(\sReprScale, \sReprScale)^T$, $\sReprScale\geq0$. One example of a maximum \sReprCCS{} ray appears in Fig.~\ref{fig:ct-path}.

We now state a key theoretical proposition of this section, showing that it suffices to search for the optimal solution on a specific triangle of the \tCCS
\proofat{sec:proof-sufficient-search-triangle}.
\begin{proposition}
  \label{thm:sufficient-search-triangle}
  The maximum of the controlled impact function $\sffitCCS{}$ is attained at a point which lies in the (filled) triangle $\sSST\eqdef\{(\sPath[1]*,\sPath[1]*),(\sPath[2]*,\sPath[2]*),\sPath*\}$, with vertices the optimal c-t point $\sPath*=(\sPath[1]*,\sPath[2]*)$ and its horizontal and vertical projections onto the maximum $\sReprCCS{}$ ray. We call this region the \temph{\acl{SST}}.
\end{proposition}

The sufficiency of the \tSST reduces the search space by at least half, which happens in the worst case scenario that the optimal c-t point $\sPath*$ is on the North-West or South-East points. More importantly, we can efficiently optimise $\sffitCCS{}$ along specific directions within this region.

We now describe these directions. For each ordinate $\siSub_2\in 0,\ldots,\fnSet[2]\sSub$ let the (West-to-East) \tdef{horizontal sequence} be
\begin{align}
  \sHSeq[\siSub_2]{}\eqdef\big(\sHSeq[\siSub_2]0,\ldots,\sHSeq[\siSub_2]{{\fnSet[1]\sSub}}\big) 
  =\big((0,\siSub_2),\ldots,(\fnSet[1]\sSub,\siSub_2)\big)\;.
\end{align}
Similarly, for each abscissa $\siSub_1\in 0,\ldots,\fnSet[1]\sSub$ we define the (South-to-North) \tdef{vertical sequence}
\begin{align}
  \sVSeq[\siSub_1]{}\eqdef\big(\sVSeq[\siSub_1]0,\ldots,\sVSeq[\siSub_1]{{\fnSet[2]\sSub}}\big)
  =\big((\siSub_1,0),\ldots,(\siSub_1,\fnSet[2]\sSub)\big)\;.
\end{align}
When the transformed controlled impact function $\sffitCCS\sPInd$ is computed along the elements of certain of those sequences, it forms concave sequences, as we show below  \proofat{sec:proof-sequence-concavity}, with the direct implication that the maximal value of $\sffit$ along these sequences can be computed in $O(\log n)$.

\begin{proposition}
  \label{thm:sequence-concavity}
  Consider the values of the controlled impact function $\sffitCCS{}$ as they are computed along a horizontal sequence $\sHSeq{}$; these form the sequence $(\sffitCCS{}\circ\sHSeq{})(\siCnt)$, which for $\siCnt\leq\siSub_2$ is a concave sequence preceding the maximum $\sReprCCS{}$ ray. Similarly, $(\sffitCCS{}\circ\sVSeq{})(\siCnt)$ is a concave sequence for $\siCnt\leq\siSub_1$. 
\end{proposition}
Observing the example of the concave sequences of $\sffitCCS{}$ along the horizontal and vertical directions shown in Fig.~\ref{fig:ct-path}, we notice that we can cover the entire \tSST with an appropriate selection of these concave sequences. This allows for an efficient optimisation procedure requiring $O(n\log n)$ time, which is described in Algorithm~\ref{alg:fixed} and operates as follows.

\begin{algorithm}
  \caption{\acl{RAWR}}\label{alg:fixed}
  \KwIn{Population $\sPop$ (sorted w.r.t $\svtar$, descending)}
  \KwIn{Subgroup $\sSub$}
  \KwOut{Tight optimistic estimator $\sfoest$ of Eq.~\eqref{eq:objective-ours}}
  $\sPath* \gets \fTSearch{\sCovCentCCS{}\circ\sPath[]{}}1{|\sSub|}$\;\label{ln:ct-path-opt}

  $\siAlgBeg\gets \min\{\sPath[1]*,\sPath[2]*\}$\;
  \uIf{$\sPath[1]*<\sPath[2]*$}{
\label{ln:ct-above}
    $\siAlgEnd\gets \min\{\sPath[2]*,\fnSet[2]\sSub\}$\;
    \For{$\siAlg$ from $\siAlgBeg$ to $\siAlgEnd$}{
\label{ln:horiz-start}
$\phi \gets \fTSearch{\sffitCCS{}\!\circ\sHSeq[\siAlg]{}}{\siAlgBeg}{\siAlgEnd}$\;\label{ln:search-horizontal}

      $\sfoest \gets \max\{\sfoest,\phi\}$\;
\label{ln:horiz-end}
    }
  }
  \Else{
\label{ln:ct-below}
    $\siAlgEnd\gets \min\{\sPath[1]*,\fnSet[1]\sSub\}$\;\label{ln:vert-start}
    \For{$\siAlg$ from $\siAlgBeg$ to $\siAlgEnd$}{
      $\phi \gets \fTSearch{\sffitCCS{}\!\circ\sVSeq[\siAlg]{}}{\siAlgBeg}{\siAlgEnd}$\;
      $\sfoest \gets \max\{\sfoest,\phi\}$\;
\label{ln:vert-end}
    }
  }
  \Return{$\sfoest$}\;
\end{algorithm}

First, the optimal c-t point $\sPath*$ is computed in $O(\log n)$ time, along the concave sequence $\sPath{}$ (line~\ref{ln:ct-path-opt}); this point locates the \tSST.\@ 
If $\sPath*$ lies above the maximum $\sReprCCS{}$ ray (line~\ref{ln:ct-above}-\ref{ln:horiz-end}), the points of $\sSST$ lie along horizontal sub-sequences preceding the maximum $\sReprCCS{}$ ray; the $\sffitCCS{}$ values along each of them form a concave sequence, whose maximum can be found in $O(\log n)$ (ln.~\ref{ln:search-horizontal}). There are at most $\fnSet[2]\sSub$ such directions in $\sSST$, and they can all be scanned (ln.~\ref{ln:horiz-start}-\ref{ln:horiz-end}) in a total of $O(n\log n)$ time. Similarly, when $\sPath*$ lies below the maximum $\sReprCCS{}$ line (ln.~\ref{ln:ct-below}), we optimise along the vertical sub-sequences (ln.~\ref{ln:vert-start}-\ref{ln:vert-end}).\@

\section{Related Work}\label{sec:related}

Whereas subgroup discovery~\cite{klosgen1996explora} is well-studied in general~\cite{atzmueller_subgroup_2015,wrobel1997algorithm,boley_non-redundant_2009, grosskreutz_tight_2008, lemmerich2016fast, duivesteijn2016exceptional}, to the best of our knowledge the notion of representativeness beyond coverage has not been studied in depth. 

In pattern mining in general, there has been ample attention to iteratively discovering patterns that are surprising given background knowledge, for example with regard to permutation testing~\cite{gionis:07:assessing,hanhijarvi:09:tell}, or a maximum entropy distribution~\cite{tatti:08:signifsets,konto:10:sdm,debie:11:dami,mampaey:12:mtv}. While seemingly related, representativeness is not guaranteed by unexpectedness: adding a pattern $X$ to our background knowledge does not ensure that, relative to $X$, the now-most-surprising pattern will be representative with regard to either pattern $X$, or to the whole population.

Another seemingly obvious relation that turns out to be much more subtle is that to fairness in classification. A truly representative pattern implies statistical parity with regard to the control variable, although it is worth noting that both Dwork and Mullainathan explicitly mention that statistical parity should not be equated with fairness, as it can potentially be ``blatantly unfair'' on an individual level~\cite{zemel_learning_2013,kleinberg_inherent_2016}.

In recent work, Feldman et al.~\cite{feldman2015certifying} studied the notion of ``disparate impact''---a legal term that says that the probability ratio of treatment (e.g., job offer) for the different groups must be at least 0.8---and proposed as a general technique to remove disparate impact via data pre-processing. In other words, unlike our approach, the global distribution is changed to de-correlate sensitive and target attributes. Related as it may be, their work clearly fails to extend to local pattern mining, as in the latter, it does not suffice to model the global distribution.

Perhaps closest to our approach is the line of work by Calders et al.~\cite{calders_why_2013}, who studied the goal of achieving statistical parity in classification with different methods, including naive bayes~\cite{calders_three_2010} and decision trees~\cite{kamiran_discrimination_2010}. Kamishima et al.~\cite{kamishima_fairness-aware_2012} consider a form of fairness that is related to statistical parity, although implicitly: during logistic regression a regularisation term is used, measuring the KL divergence between sensitive attribute and prediction. Although related, it is unclear whether these methods can be utilised in the highly demanding \tBNB search, typically able to optimise over exponentially-sized discrete spaces of arbitrary subgroup descriptor languages.

Closest in terms of pattern mining, but relatively distant with regard to statistical parity, is the work by Pedreschi et al.~\cite{pedreshi_discrimination-aware_2008} on discrimination-aware pattern mining. Instead of subgroups, the authors focus on mining association rules that may only include a sensitive item if this does not improve the confidence of the rule by more than $\alpha$.

\section{Experiments}
\label{sec:exps}

\begin{table}
    \centering
    \resizebox{\linewidth}{!}{
      \begin{tabu}{llllrr}
        \toprule
        Dataset& Target \svtar& Control \svctr& $\sappr$ & $|\sVars|$ & $|\sPop|$ \\
        \midrule
 baseball & Salary & Fr.Ag.Elig. & 1.0 &  16 & 268 \\ 
  gold & Evdw-Evdw0 & Odd & 1.0 &  19 & 12200 \\ 
  homicide & Victims & PerpRace & 1.0 &  10 & 47236 \\ 
  students & G3 & failures & 0.5 &  31 & 366 \\ 
  wine & quality & colour & 1.0 &  12 & 3198 \\ 
  \midrule
 abalone & Rings & Height & 1.0 &   8 & 4144 \\ 
  ailerons & Sa & RollRate & 1.0 &   5 & 7108 \\ 
  airfoil & NoiseLevel & Displacement & 1.0 &   5 & 1480 \\ 
  autompg & Mpg & Cylinders & 1.0 &   8 & 380 \\ 
  bike & registered & atemp & 1.0 &  13 & 730 \\ 
  california & Med.Value & Latitude & 0.5 &   8 & 20502 \\ 
  compactiv & usr & freeswap & 0.7 &  21 & 8192 \\ 
  concrete & Strength & Age & 1.0 &   8 & 562 \\ 
  elevators & Goal & DiffRollRate & 0.3 &  18 & 16020 \\ 
  forestfires & Area & Month & 0.6 &  12 & 394 \\ 
  house & Price & P14p9 & 0.3 &  16 & 22784 \\ 
  mortgage & 30YRate & Mat.Rate3Y & 0.8 &  15 & 1044 \\ 
  mv & Y & X6 & 0.9 &  10 & 40768 \\ 
  pole & Output & Att0 & 0.3 &  26 & 14586 \\ 
  puma32h & thetadd6 & theta5 & 0.7 &  32 & 8192 \\ 
  stock & Company10 & Company4 & 1.0 &   9 & 950 \\ 
  treasury & X1Rate & CMat.Rate3Y & 0.4 &  15 & 1044 \\ 
  wankara & AvgTmp & MaxTemp & 0.6 &   9 & 318 \\ 
  wizmir & AvgTmp & MaxTemp & 0.5 &   9 & 1458 \\

        \bottomrule
    \end{tabu}
    }
    \caption{Used datasets, for qualitative (top) and quantitative (bottom) analysis. Listed are the number of attributes $|\sVars|$ and number of rows $|\sPop|$, as well as running configurations: target and control variables, and approximation factor $\sappr$. The latter is decreased by $0.1$ every time \tBRIG exceeds a timeout of 6 hours, or terminates due to exceeding $256$GB of memory.}\label{tbl:dataset-specs}
\end{table}

In this section we evaluate our extended impact function $\sffit$, as well as \tRAWR, our implementation of its tight optimistic estimator for use by the \tBNB\ algorithm. We provide qualitative and quantitative demonstrations of superior representativeness in the discovered subgroups, and we also report runtime measurements on a variety of datasets.

For these tasks we implemented\footnote{
Our source code is available within the realKD tool \url{bitbucket.org/realKD/}.
} both \tRAWR\ and the non-tight, representativeness oblivious \tBRIG, which we use as a baseline. We then run the \tBNB\ algorithm, using each of them.

\subsection{Mining more representative results}
We now assess qualitatively and quantitatively the representativeness of the discovered subgroups, for different values of the $\sweight$ parameter. We first study $5$ datasets (retrieved from the UCI ML repository\cite{uciml} and the Murder Accountability Project \url{http://www.murderdata.org/}) which contain intuitively interpretable controls (Table~\ref{tbl:dataset-specs}, top). To rule out the effect of unbalanced classes, and for our algorithm to be applicable, we stratify the datasets over the control classes; we then perform subgroup discovery and measure the $\sRepr$ and $\sCovCent$ scores of the discovered subgroups, as we increase the $\sweight$ parameter (Fig.~\ref{fig:weights}).

Obviously, a value of $\sweight=0$ corresponds to the representativeness-oblivious impact function $\sCovCent$ of Eq.~\eqref{eq:objective-other}. Depending on the dataset and choice of $\svtar$ and $\svctr$, the discovered subgroups for this case may be representative, although this is not guaranteed. However, as the $\sweight$ parameter increases, the added $\sRepr$ factor comes in effect, yielding consistently more representative subgroups (Figure~\ref{fig:weight-representativeness}). As expected, the $\sCovCent$ score may drop, demonstrating that $\sweight$ controls the trade-off between the two factors (Figure~\ref{fig:weight-covcent}). At the same time, it is guaranteed that no score can be increased without the decrease of the other, by choosing a subgroup other than the discovered.

We next delve into the subgroups discovered in selected datasets. We first focus on the \tDS{Homicide} dataset, which tracks homicide cases, matched with background data on perpetrators and their victims, alongside the number of victims per case. We use the latter as a target variable to measure violence and seek to gain insight on attributes leading to increased violence, as captured by binary control variables. For each studied variable, we stratify the dataset and report the discovered subgroups (Table~\ref{tbl:subgroups}), for increasing $\sweight$ parameter.

We first consider the effect of the \tvar{Perpetrator Sex}. The subgroup discovered without the $\sRepr$ term rediscovers the unsurprising fact that males are more violent than females. To uncover further potentially underlying factors, we use the \tvar{Perpetrator Sex} as a control variable and perform subgroup discovery using the controlled impact function. As $\sweight$ parameter increases, the discovered subgroups hold for both male and female perpetrators, leading to the discovery that \temph{Caucasian victims attract more violence}, and further that \temph{no sex is more violent when it comes to older victims}.

In another example, we study an application for fair subgroup discovery. Consider that a baseball team decides to increase its players salary and seeks to find the factors that lead to higher income drawing experience from other team managements. At the same time, the raise should not be unfavourable to players who are contractually bound to one particular team, in contrast to the Free Agent eligible players, which might earn more lest they leave the team. Using the \tvar{FreeAgencyEligibility} variable as control, more objective criteria describing high salaries are discovered.

\subsection{Evaluating the performance of the proposed tight estimator}
\begin{figure}
  \newcommand{\plotData}[2][]{
    \pgfplotstableread{#2}{\table}
    \pgfplotstablegetcolsof{#2}
    \pgfmathtruncatemacro\numberofcols{\pgfplotsretval-1}
    \pgfplotsinvokeforeach{1,...,\numberofcols}{
      \pgfplotstablegetcolumnnamebyindex{##1}\of{\table}\to{\colname}
      \addplot table [y index=##1,#1] {#2}; 
      \addlegendentryexpanded{\colname}
    }
  }
  \pgfplotsset{
    eda axis/.style={
      ticklabel shift = 0,
      scaled ticks = false,
      x label style = {},
      y label style = {normal shift={y}{-5pt}},
      xtick style = {font=\footnotesize},
      ytick style = {font=\footnotesize},
      x tick label style = {yshift=-2pt},
      y tick label style = {xshift=-2pt},
      xtick style = {normal shift={x}{-7pt}},
      ytick style = {normal shift={y}{-7pt}},
      axis lines*=left,
      x label style={normal shift={x}{-9pt}},
      y label style={normal shift={y}{-9pt}},
      x axis line style = {thick,normal shift={x}{-8pt}},
      y axis line style = {thick,normal shift={y}{-8pt}},
      enlargelimits = false,
    },
  }
  \begin{minipage}[t]{0.5\linewidth}%
    \begin{tikzpicture}
      \begin{axis}[
        eda line,
        eda axis,
        ymax = 1.0, ymin = 0.0,
        ylabel={Representativeness {$\sRepr$}},
        xlabel={$\sweight$},
        legend pos=south east,
        height=3.8cm,
        width=\textwidth,
        legend style={nodes={scale=0.92, transform shape}, at={(1.0,0.12)},anchor=south east}, legend columns=2]
        \plotData{./data/scores-r-special.dat}
      \end{axis}
    \end{tikzpicture}%
    \vspace{-1.5ex}%
    \subcaption{Representativenes score $\sRepr$.}\label{fig:weight-representativeness}
  \end{minipage}%
  \begin{minipage}[t]{0.5\linewidth}%
    \begin{tikzpicture}
      \begin{axis}[
        eda line,
        eda axis,
        y label style = {normal shift={y}{-3pt}},
        ymax = 0.15, ymin = 0.0,
        ylabel={Coverage-Tendency $\sCovCent$},
        xlabel={$\sweight$},
        legend pos=south east,
        height=3.8cm,
        width=\textwidth,
        scaled y ticks={base 10:2},
        ]
        \plotData{./data/scores-ct-special.dat}
        \legend{}
      \end{axis}
    \end{tikzpicture}%
    \vspace{-1.5ex}%
    \subcaption{Coverage-tendency score $\sCovCent$.}\label{fig:weight-covcent}
  \end{minipage}
  \caption{Scores of subgroups discovered in the qualitative datasets (Table~\ref{tbl:dataset-specs}, top). Tuning $\sweight$ effectively controls the trade-off between Representativeness and coverage-tendency.}\label{fig:weights}
\end{figure}
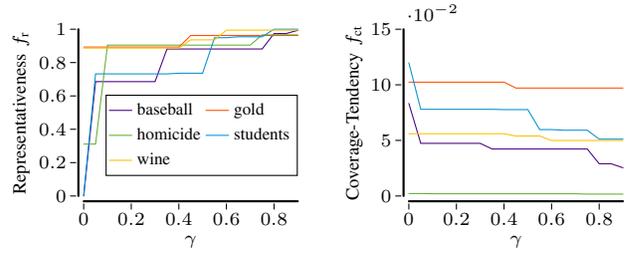

\begin{table}
  \centering
  \resizebox{0.96\linewidth}{!}{
    \begin{tabu}{lrrrrrr}
      \toprule
         \multirow{2}{*}{Dataset}
            & \multicolumn{2}{c}{$\sweight=0.4$}
            & \multicolumn{2}{c}{$\sweight=0.5$}
            & \multicolumn{2}{c}{$\sweight=0.6$}
            \\\cmidrule{2-3}\cmidrule{4-5}\cmidrule{6-7}
         & \tRAWR&\tBRIG
         & \tRAWR&\tBRIG
         & \tRAWR&\tBRIG\\\midrule
 gold & \tnopt{172} & \topt{101} & \tnopt{210} & \topt{99} & \tnopt{224} & \topt{121} \\ 
  wine & \tnopt{296} & \topt{267} & \tnopt{349} & \topt{305} & \tnopt{375} & \topt{360} \\ 
  house & \tnopt{14} & \topt{5} & \tnopt{13} & \topt{5} & \tnopt{17} & \topt{4} \\ 
  stock & \tnopt{15} & \topt{8} & \tnopt{16} & \topt{10} & \tnopt{17} & \topt{12} \\ 
  california & \tnopt{3} & \topt{2} & \tnopt{4} & \topt{2} & \tnopt{4} & \topt{2} \\ 
  pole & \tnopt{4} & \topt{2} & \tnopt{3} & \topt{2} & \tnopt{3} & \topt{1} \\ 
  airfoil & \topt{9} & \tnopt{9} & \topt{7} & \tnopt{8} & \tnopt{9} & \topt{9} \\ 
  concrete & \topt{4} & \tnopt{6} & \topt{5} & \tnopt{5} & \topt{6} & \tnopt{6} \\ 
  elevators & \tnopt{3} & \topt{3} & \tnopt{3} & \topt{3} & \topt{2} & \tnopt{2} \\ 
  puma32h & \topt{78} & \tnopt{84} & \topt{82} & \tnopt{83} & \topt{83} & \tnopt{85} \\ 
  baseball & \topt{20} & \tnopt{22} & \topt{17} & \tnopt{21} & \topt{16} & \tnopt{19} \\ 
  bike & \topt{5} & \tnopt{10} & \topt{5} & \tnopt{12} & \topt{6} & \tnopt{11} \\ 
  forestfires & \topt{184} & \tnopt{272} & \topt{178} & \tnopt{186} & \topt{193} & \tnopt{217} \\ 
  homicide & \topt{154} & \tnopt{219} & \topt{171} & \tnopt{247} & \topt{169} & \tnopt{236} \\ 
  ailerons & \topt{206} & \tnopt{317} & \topt{297} & \tnopt{486} & \topt{703} & \tnopt{797} \\ 
  compactiv & \tnopt{504} & \topt{450} & \tnopt{442} & \topt{349} & \topt{1299} & \tnopt{3397} \\ 
  mv & \topt{3877} & \tnopt{6837} & \topt{3243} & \tnopt{5273} & \topt{2300} & \tnopt{5026} \\ 
  students & \topt{19} & \tnopt{175} & \topt{63} & \tnopt{2638} & \topt{126} & \tnopt{4768} \\ 
  autompg & \topt{6} & \tnopt{3229} & \topt{6} & \tnopt{5577} & \topt{11} & \tnopt{10591} \\ 
  abalone & \topt{869} & \tnopt{1307} & \topt{1883} & \tnopt{3876} & \topt{3639} & \tnopt{17575} \\ 
  wankara & \topt{1} & \tnopt{116} & \topt{1} & \tnopt{2543} & \topt{1} & \tnopt{13977} \\ 
  mortgage & \topt{56} & \tmis & \topt{198} & \tmis & \topt{12568} & \tmis \\ 
  treasury & \topt{1} & \topt{1} & \topt{1} & \topt{1} & \topt{1} & \tmis \\ 
  wizmir & \tnopt{5} & \topt{3} & \topt{2} & \tnopt{1648} & \topt{3} & \tmis \\

         \bottomrule
     \end{tabu}
     }
  \caption{[Bold is better] Runtime comparison of \tRAWR and \tBRIG over different $\sweight$ parameters for all datasets, sorted in increasing time difference. Using \tBRIG on the last 3 datasets exceeds our $256$GB of memory, so results are not available.}\label{tbl:runtimes}
\end{table}

We now evaluate the performance of the \tRAWR\ implementation. To sample a broad variety of datasets, we used all of the regression datasets from the KEEL database~\cite{alcala-fdez_keel_2011} with a number of variables $8\leq |\sVars| \leq 40$ (Table~\ref{tbl:dataset-specs}, bottom). As a target variable we used the designated regression variable. To emulate a purported scenario of controlling for the main data trend, we use as control the first variable that appears in the subgroup descriptor discovered for $\sweight=0$; if this variable is real we discretise it around the median. Next, we stratify the dataset on the control variable. We start with an approximation factor of $\sappr=1$, corresponding to exact computation; when all \tBRIG\ invocations for a dataset fail, due to either a runtime of more than 6 hours or exceeding $256$GB of available memory, we decrease $\sappr$ by 0.1 and repeat.

We assess the performance of our algorithm w.r.t.\ the number of searched nodes during each \tBNB\ invocation and also the needed runtime (Fig.~\ref{fig:performance}); we set $\sweight=0.6$, corresponding to a reasonably practical scenario. As our proposed optimistic estimator is tighter, it is yielding a significantly better pruning performance. What is more, our implementation seems to make use of the better pruning achieved, in order to attain running times that are comparable to those of \tBRIG, or in some cases up to 4 orders of magnitude better. Further numerical results are reported in Table~\ref{tbl:runtimes}, for a set of sensible weights $\sweight\in\{0.4,0.5,0.6\}$. These show a superiority of our estimator especially for higher $\sweight$ values, where \tBRIG\ is less tight.

Furthermore, note that the number of nodes is a key factor contributing to the memory requirement of the \tBNB\ algorithm. As such, even for dataset on which the computation time of these implementations might be comparable, it is sometimes the case that the decreased number of nodes is enabling the computation using \tRAWR, where otherwise \tBRIG\ would exceed available memory, e.g., in the last $3$ datasets of Table~\ref{tbl:runtimes}. 

\begin{table}
\begin{tabular}{@{}cr@{}r@{,}r@{}l@{\enspace}c@{\enspace}r@{\enspace}r@{}}
  \toprule
  \multirow{13}{*}{\rotatebox{90}{\tDS{homicide}}}
  &\multicolumn{4}{c}{$\sweight$}& Subgroup describing $\sSub$& $\sRepr(\sSub)$& $\sCovCent(\sSub)$\\
 \midrule
 &\multicolumn{7}{c}{Control: \tvar{Perpetrator Sex}}\\\cmidrule{2-8}
    &$($&$0$&$0.09$&$]$& Crime=Murder, Vict.=White, Perp=\tmale & 0.00 & 0.002 \\ 
    &$($&$0.09$&$0.75$&$]$& Vict.=White & 0.89 & 0.002 \\ 
  &$[$&$0.75$&$\ldots$&$)$& Vict.Age= $\neg$V.Lo, Vict.=White & 0.99 & 0.001\\
  \cmidrule{2-8}
 &\multicolumn{7}{c}{Control: \tvar{Perpetrator Race}}\\\cmidrule{2-8}
    &$($&$0$&$0.6$&$]$& Crime=Murder, Vict=\tfemale, Perp.= $\neg$V.Old & 0.90 & 0.002 \\ 
  &$[$&$0.6$&$\ldots$&$)$& Crime=Murder, Vict=\tfemale, Perp.= $\neg$Old & 0.98 & 0.002 \\
  \midrule
  \multirow{5}{*}{\rotatebox{90}{\tDS{baseball}}}&\multicolumn{7}{c}{Control: \tvar{Free Agent Eligibility}}\\\cmidrule{2-8}
    &$($&$0$&$0.09$&$]$& OnBase= $\neg$V.Lo, F.A.Eligible= \tyes & 0.00 & 0.083 \\ 
    &$($&$0.09$&$0.33$&$]$& OnBase=HI & 0.69 & 0.047 \\ 
    &$($&$0.33$&$0.8$&$]$& Batting= $\neg$V.Lo, OnBase= $\neg$Lo & 0.88 & 0.042 \\ 
  &$[$&$0.8$&$\ldots$&$)$& Batting=$\neg$V.Lo, OnBase= $\neg$Lo, Fr.Ag.= \tno & 0.97 & 0.029 \\
  \bottomrule
\end{tabular}

    \caption{Discovered subgroups for a varying $\sweight$ parameter, for the datasets \tDS{homicide} (above) and \tDS{baseball} (below). Increasing $\sweight$ produces more representative subgroups.}\label{tbl:subgroups}
\end{table}

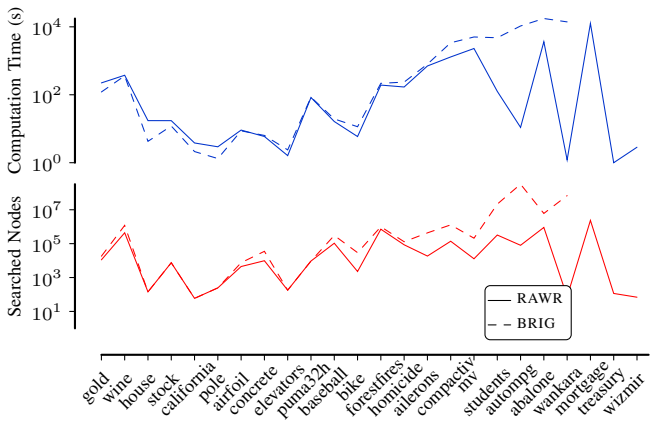
\begin{figure}
  \pgfplotstableread[header=true]{./data/timings-keep.dat}\tbltimings
  \pgfplotstablegetrowsof{\tbltimings}
  \pgfmathsetmacro{\xmax}{\pgfplotsretval-0.5}
  \tikzset{
    brig line/.style={dashed},
    rawr line/.style={solid},
    time line/.style={green!20!blue},
    nodes line/.style={red},
  }
  \begin{tikzpicture}[
    ]
    \pgfplotsset{
      width=\linewidth,
      height=3.5cm,
      grid style={dashed,gray},
    }
    
    \begin{axis}[
      eda line,
      ylabel={\strut Computation Time (s)},
      scaled y ticks={base 10:-5},
      ymode=log,
      axis x line=none,
      xmax=\xmax,
      x axis line style={draw opacity=0},
      x tick label style={rotate=45,text width=2em,anchor=north east,align=right,scale=1,font=\scriptsize},
      xmajorgrids=true,
      separate axis lines,
      ]
      \addplot[rawr line,time line,mark=o] table[x expr=\coordindex,y={060RAWRTimeVal}]{\tbltimings};
      \addplot[brig line,time line,mark=o] table[x expr=\coordindex,y={060BRIGTimeVal}]{\tbltimings};
    \end{axis}
  \end{tikzpicture}\\%
  \begin{tikzpicture}[
    ]
    \pgfplotsset{
      width=\linewidth,
      height=3.5cm,
    }
    \begin{axis}[
      eda line,
      ylabel={\strut Searched Nodes},
      xticklabels from table={\tbltimings}{Dataset},
      xtick=data,
      scaled y ticks={base 10:-5},
      ymode=log,
      xmax=\xmax,
      ytickten={1,3,5,7},
      ymin=0,
      x tick label style={rotate=45,text width=2em,anchor=north east,align=right,scale=1,font=\scriptsize},
      unbounded coords=jump,
      legend style={
        at={(0.7,0.3)},
        legend columns=1,
        rounded corners=2pt,
        anchor=north west},
      ]
      \addlegendimage{rawr line}\addlegendentry{\acs{RAWR}}
      \addlegendimage{brig line}\addlegendentry{\acs{BRIG}}
      \addplot[draw=none,forget plot] table[x expr=\coordindex,y={060RAWRTimeVal}]{\tbltimings};
      \addplot[brig line,nodes line] table[x expr=\coordindex,y={060BRIGNodesVal}]{\tbltimings};
      \addplot[rawr line,nodes line] table[x expr=\coordindex,y={060RAWRNodesVal}]{\tbltimings};
    \end{axis}
  \end{tikzpicture}
  \caption{[Lower is better] Performance comparison of \tRAWR\ (solid) and \tBRIG\ (dashed), for runtime (blue) and searched nodes (red), with $\sweight=0.6$. The datasets (x axis) are sorted in increasing time difference. Using \tBRIG on the last 3 datasets exceeds our $256$GB of memory, so results are not available.}\label{fig:performance}
\end{figure}

\section{Discussion}
\label{sec:discussion}

Our introduced method guarantees the optimality of the results given the specified parameter, while optionally enabling a faster computation by relaxing the optimality guarantee.

The sole parameter $\sweight$ of our method remains intuitive in its interpretation and possibly in its selection, regardless of the input, with the zero value corresponding to a vanishing effect of our extension and a high value an increased weight of it. Nonetheless, not every dataset is equally sensitive to the intermediate values and the researcher is still required to make educated guesses based equally well on expert knowledge or resort to a trial and error scheme.

A further point of interest is the rate of convergence. Inheriting the weak points of the \tBNB algorithm, the worst case complexity is no better than exponential, although in no real world cases has this been observed. Additionally, there seems to be no good estimate on the difficulty of the dataset.

As a downside, our implementation of the tight estimator for our objective function requires a balanced, binary dataset. However, the case of binary classes is amongst the most usual, and the former can be solved by stratification as a workaround.

In the future we are planing to further broaden the application settings of our estimator to more than binary classes and non-balanced datasets, as well as to investigate the underlying causes of this inherently difficult optimisation problem.

\apxtrue
\section{Conclusion}\label{sec:conclusion}

We introduce the problem of representativeness aware subgroup discovery, where we want to discover subgroups that are exceptional with regard to the target variable, yet at the same time be to statistical parity with regard to the control variable. We show how we can achieve this by extending the typically used impact function by incorporating a tuneable representativeness factor. We propose a tight optimistic estimator for the newly representative aware impact function, and give an efficient algorithm to compute it in $O(n \log n)$ time. Experiments show it may lead to orders of magnitude fewer node expansions, compared to the representative ignorant estimator, which is leveraged in similar orders of speedup.
In future work we aim to extend our theory to  nominal control variables, and studying exceptional representative model mining~\cite{duivesteijn2016exceptional}.

\providecommand{\noopsort}[1]{}

\ifapx
\appendix
\section{Appendix} \label{sec:appendix}
In this section we provide proofs for our theoretical claims.
\subsection{Proof of Lemma~\ref{thm:optimal-ct-path}: optimal c-t properties}\label{sec:proof-optimal-ct-path}
Let $\sSub_\siCnt^*\subseteq\sSub$ be the set attaining the highest $\sCovCent$ value among those with cardinality $\siCnt$. We now reinterpret Eq.~\eqref{eq:ct-path-definition} as follows: the element $\sPath\siCnt$ is equal to the index of the equi-count refinement set $\sSelPar[\sPath*]$ containing $\sSub_\siCnt^*$.
Within all sets with a fixed cardinality $\sCov$ remains constant, and $\sSub_\siCnt^*$ is the set with the maximal central tendency $\sCent$; we can then show, similar to Proposition~\ref{thm:efficient-naive-opt}, that the maximiser of $\sCent$  contains the topmost $\siCnt$ target values. Altogether, $\sPath\siCnt$ is exactly the class count of
\begin{align}
  \sSub_\siCnt^*\eqdef
  \argmax_{|\sSel|=\siCnt,\;\sSel\subseteq\sSub}\sCovCent(\sSel) =
  \argmax_{|\sSel|=\siCnt,\;\sSel\subseteq\sSub}\sCent(\sSel) =
  \bigcup_{i=1}^\siCnt\{\sitem_i\}\;,
\end{align}
whose control class count is equal to the quantity in Eq~\eqref{eq:ct-path-solution}. 

  To show (ii) we proceed as follows. Since $\sSub^*_\siCnt$ contains the top-$\siCnt$ elements, we rewrite Eq.~\eqref{eq:objective-covcent-cspace} as
  \begin{align}
    (\sCovCent\circ \sPath{})(\siCnt) = \sCovCent(\sPath{\siCnt}) = & \sconstt \sum_{i=1}^\siCnt\svtar_i -\sconstc\siCnt\;,
  \end{align}
  with discrete derivatives
  \begin{align}
    \Delta_\siCnt(\sCovCent\circ \sPath{}) =& \sCovCent(\sPath{\siCnt+1})-\sCovCent(\sPath\siCnt)=
                                             \sconstt\svtar_{\siCnt+1}-\sconstc\\
    \Delta^2_\siCnt(\sCovCent\circ \sPath{})
    =& \sconstt(\svtar_{\siCnt+1}-\svtar_\siCnt) \leq 0\;,
  \end{align}
  where the last inequality holds because $\svtar_\siCnt$ are decreasing. The negativity of the second discrete derivative, shows the concavity of the sequence.
\subsection{Proof of Proposition~\ref{thm:sufficient-search-triangle}: \acl*{SST}}\label{sec:proof-sufficient-search-triangle}

 This proof involves partitioning the \tCCS in compact regions surrounding the \tSST, within which the monotonicity of the factors $\sCovCentCCS{}$ and $\sReprCCS{}$ remains constant, when computed along any horizontal or vertical sequences. All the sequences formed in this way increase toward the region boundary intersecting with the \tSST, and so no maximiser of $\sffitCCS{}$ can lie within these regions, except on the intersection of the region and the \tSST.

 To show the above, we study both factors, starting with $\sCovCentCCS{}$.
 \begin{lemma}[Domination of the $\sCovCentCCS{}$ factor]\label{thm:coverage-tendency-domination}
   The impact value computed along any horizontal sequence is increasing until the abscissa $\sPath[1]*$ of the optimal c-t point, and decreasing afterwards. Similarly, the impact value computed along any vertical sequence is increasing until the ordinal $\sPath[2]*$ of the optimal c-t point, and decreasing afterwards. Formally, 
  \begin{align}
    \label{eq:ct-domination-inequality-left}
    \sCovCentCCS{\sPInd+\sSBVec}&\geq \sCovCentCCS\sPInd\;, &\sPInd[\siCls]< \sPath[\siCls]*\\
    \label{eq:ct-domination-inequality-right}
    \sCovCentCCS{\sPInd+\sSBVec}&\leq \sCovCentCCS\sPInd\;, &\sPInd[\siCls]\geq \sPath[\siCls]*
  \end{align}
  \proof
  Denote the \tdef{optimal c-t path index} $\sPathOptInd$ to be the index within the c-t path sequence attaining the maximum c-t value, $\sPath{\sPathOptInd}=\sPath*$, so that
\begin{align}
    \label{eq:ct-path-value-inequalities}
    \begin{split}
      \sCovCent(\sPath{\siCnt+1}) \geq \sCovCent(\sPath\siCnt) &\qquad \siCnt < \sPathOptInd\\
      \sCovCent(\sPath{\siCnt+1}) \leq \sCovCent(\sPath\siCnt) &\qquad \siCnt \geq \sPathOptInd
      \end{split}\enspace,
\end{align}
due to the concavity of the sequence $(\sffitCCS{}\circ\sPath{})(\siCnt)$.

 However, for any two consecutive points on the path, we can compute
$ \sCovCent(\sPath{\siCnt+1}) - \sCovCent(\sPath\siCnt) =
      \sconstt \svtar_{\siCnt+1}-\sconstc$, which combined with Eq.~\eqref{eq:ct-path-value-inequalities} yields
  \begin{align}
    \label{eq:ct-path-value-difference-inequalities}
    \begin{split}
      \sconstt \svtar_{\siCnt+1}-\sconstc\geq 0&\qquad \siCnt < \sPathOptInd\\
      \sconstt \svtar_{\siCnt+1}-\sconstc\leq 0&\qquad \siCnt \geq \sPathOptInd
      \end{split}\enspace.
  \end{align}
  
Moreover, using Eq.~\eqref{eq:objective-covcent-cspace} we can express the $\sCovCentCCS{}$ value of the point next to $\sPInd$ in \tCCS along dimension $\siCls$ as
  \begin{align}
    \sCovCentCCS{\sPInd+\sSBVec} = \sconstt
    \sum_{\siCls=1}^2
    \left(
    \sum_{\siSub=1}^{\sPInd[\siCls]}\sfvtar{\siSub} +
    \sfvtar{\sPInd[\siCls]+1}
    \right)
    - \sconstc(\sPInd[1]+\sPInd[2])\;,
  \end{align}
and so the difference between the $\sCovCentCCS{}$ values of these neighbouring points becomes
  \begin{align}
    \label{eq:ct-value-difference-ccs}
    \sCovCentCCS{\sPInd+\sSBVec} - \sCovCentCCS\sPInd = \sconstt
    \sfvtar{\sPInd[\siCls]+1}
    - \sconstc\enspace,
  \end{align}
  which is the quantity whose sign we study.
  According to Eq.~\eqref{eq:ct-path-solution}, however, $\sPath{}$ is a sequence of single step increases $\sPath{\siCnt+1}-\sPath\siCnt=\sSBVec[\sfvctr\siCnt]$, starting from the empty count $\fVec0$. In other words, the $\siCnt$-th element $\sPath\siCnt$ of the sequence increases this class count that matches the class of the item in $\sSub$ with the next greatest target value. This implies that at the optimal c-t path index $\sPathOptInd$, the optimal c-t path count $\sPath*=\sPath{\sPathOptInd}$ per class amounts exactly to the number of items with the same control class and greater target value. Moreover, for each $\sPInd[\siCls]\geq\sPath[\siCls]*$ there exists a $\siCnt\geq\sPathOptInd$ such that $\sfvtar[]\siCnt=\sfvtar[\siCls]{\sPInd[\siCls]+1}$, and similarly for each $\sPInd[\siCls]\leq\sPath[\siCls]*$ there exists a $\siCnt\leq\sPathOptInd$ such that $\sfvtar[]\siCnt=\sfvtar[\siCls]{\sPInd[\siCls]+1}$. We can now combine the two equations Eq.~\eqref{eq:ct-value-difference-ccs}, and Eq.~\eqref{eq:ct-path-value-difference-inequalities}, to show the claim of the lemma.\qed
\end{lemma}

We now proceed to show a similar behaviour of the $\sRepr$ factor.

\begin{lemma}[Total Variation domination]\label{thm:representativeness-domination}
  The composition of $\sRepr$ with every horizontal sequence $\sHSeq[\siSub]{}$, $\siSub=0,\ldots,\fnSet[1]\sSub$, and every vertical sequence $\sVSeq[\siSub]{}$, $\siSub_1=0,\ldots,\fnSet[1]\sSub$ forms the sequences $(\sRepr\circ\sHSeq[\siSub]{})(\siSeq)$ and $(\sRepr\circ\sVSeq[\siSub]{})(\siSeq)$; these are (i) uni-modal, (ii) attain a maximum at their intersection $(\siSub,\siSub)^T$ with the equi-representativeness ray $\sReprScale(1,1)^T$, $\sReprScale\geq0$, and (iii) they are concave for the indices $\siSeq=0,\ldots,\siSub$.
  \proof We first focus on the horizontal sequences $(\sRepr\circ\sHSeq[\siSub]{})(\siSeq)$ for $\siSub=0,\ldots,\fnSet[2]\sSub$ and $\siSeq=0,\ldots,\snSub[*]$. Notice that the $\sftvd(\sPInd)$ vanishes on the equi-representativeness ray $\sReprScale(1,1)^T$, that is, when $\sPInd[1]=\sPInd[2]$. Since the horizontal sequence $\sHSeq[\siSub]{}$ has a fixed ordinal of $\siSub$, the previous condition yields $\siSeq=\siSub$, which shows the correctness of (ii).

  To prove the rest of this lemma, we study the continuous analogue of $(\sRepr\circ\sHSeq[\siSub]{})(\siSeq)$
  \begin{align}
    \label{eq:tvd-continuous}
    \tilde\sRepr(\scSeq)\eqdef1-\left|\frac12-\frac\scSeq{\scSeq+\siSub}\right|\;,&& \scSeq>0\;,
  \end{align}
  which has first and second derivatives
  \begin{align}
    \label{eq:tvd-continuous-derivatives}
    \begin{split}
\tilde\sRepr'(\scSeq)&=\phantom{-2}\sign\left(\frac12-\frac\scSeq{\scSeq+\siSub}\right)\frac\siSub{(\scSeq+\siSub)^2}\\ 
\tilde\sRepr''(\scSeq)&=-2\sign\left(\frac12-\frac\scSeq{\scSeq+\siSub}\right)\frac\siSub{(\scSeq+\siSub)^3}
\end{split}&& \scSeq\neq \siSub\;.
  \end{align}
  The sign of both quantities is controlled by the sign factor, which is negative when $\scSeq<\siSub$ and positive when $\scSeq>\siSub$, and so we can reach the conclusion that $\tilde\sRepr'$ is increasing concave for $\scSeq<\siSub$ and decreasing convex for $\scSeq>\siSub$. Since $(\sRepr\circ\sVSeq[\siSub]{})(\siSeq)=\tilde\sRepr(\scSeq)$, the above properties transfer to the discrete sequence $(\sRepr\circ\sVSeq[\siSub]{})$, as well. For vertical sequences, the symmetric argumentation can be used.\qed
\end{lemma}

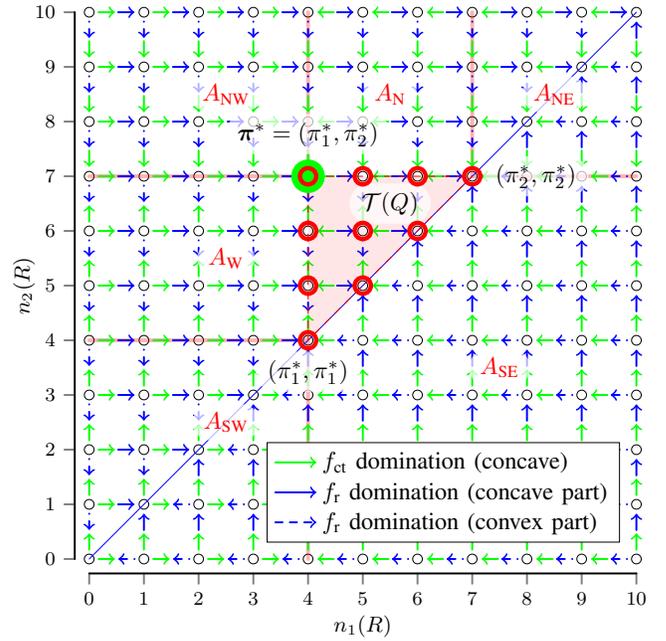
\begin{figure}
  \figWidth=1.08\linewidth
  \figHeight=1.08\linewidth
  \resizebox{\linewidth}{!}{
\begingroup
\begin{tikzpicture}
\pgfmathsetmacro\domoffx{\the\figWidth/10*\plotDomOffset/2}
\pgfmathsetmacro\domoffy{\the\figHeight/10*\plotDomOffset/2}
\pgfmathsetmacro\ds{0.5-\plotDomSize/2}
\begin{axis}[%
  small ticks,
  ccs plot 2,
  width=\figWidth,
  height=\figHeight,
  legend style={at={(axis cs:9.7,0.3)},legend cell align=left,align=left,draw=black,anchor=south east}
]
\foreach \x in {0,...,10}{
   \foreach \y in {0,...,10}{
     \edef\temp{
       \noexpand\node[draw,circle,point] at (axis cs:\x,\y) {};
     }
     \temp
   }
 }
\path[region,draw] (axis cs:0,4) -- (axis cs:4,4) -- (axis cs:4,0);
\path[region,draw] (axis cs:0,7) -- (axis cs:4,7) -- (axis cs:4,10);
\path[region,draw] (axis cs:7,10) -- (axis cs:7,7) -- (axis cs:10,7);
\foreach \i in {1,...,10}{
   \foreach \j in {1,...,\i}{
     \edef\temp{
        \noexpand\path[draw,horiz repr,convex,<-] (axis cs:\i-1+\ds,\j-1) -- (axis cs:\i-\ds,\j-1);
        \noexpand\path[draw,horiz repr,concave,->] (axis cs:\j-1+\ds,\i) -- (axis cs:\j-\ds,\i);
        \noexpand\path[draw,vert repr,convex,<-] (axis cs:\j-1,\i-1+\ds) -- (axis cs:\j-1,\i-\ds);
       	\noexpand\path[draw,vert repr,concave,->] (axis cs:\i,\j-1+\ds) -- (axis cs:\i,\j-\ds);
     }
     \temp
   }
 }
 \foreach \y in {0,...,10}{
    \foreach \x in {1,...,4}{
      \edef\temp{
        \noexpand\path[draw,horiz covcent,concave,->] (axis cs:\x-1+\ds,\y) -- (axis cs:\x-\ds,\y);
      }
      \temp
    }
    \foreach \x in {5,...,10}{
      \edef\temp{
        \noexpand\path[draw,horiz covcent,concave,<-] (axis cs:\x-1+\ds,\y) -- (axis cs:\x-\ds,\y);
      }
      \temp
    }
 }
 \foreach \x in {0,...,10}{
    \foreach \y in {1,...,7}{
      \edef\temp{
        \noexpand\path[draw,vert covcent,concave,->] (axis cs:\x,\y-1+\ds) -- (axis cs:\x,\y-\ds);
      }
      \temp
    }
    \foreach \y in {8,...,10}{
      \edef\temp{
        \noexpand\path[draw,vert covcent,concave,<-] (axis cs:\x,\y-1+\ds) -- (axis cs:\x,\y-\ds);
      }
      \temp
    }
  }
 
\path[repr,draw,thin] (axis cs:0,0) -- (axis cs:10,10);

\addplot [color=green,only marks,mark=*,ctcount,forget plot]
  table[row sep=crcr]{%
4	7\\
};

\addplot [color=red,solid,only marks,mark=o,sst,forget plot]
  table[row sep=crcr]{%
4 7\\5 7\\6 7\\7 7\\
4 6\\5 6\\6 6\\
4 5\\5 5\\
4 4\\
};

\addlegendimage{covcent,concave,->};
\addlegendentry{$\sCovCent$ domination (concave)};
\addlegendimage{repr,concave,->};
\addlegendentry{$\sRepr$ domination (concave part)};
\addlegendimage{repr,convex,->};
\addlegendentry{$\sRepr$ domination (convex part)};

\foreach \x/\y/\l in {2.5/5.5/W,2.5/8.5/NW,5.5/8.5/N,8.5/8.5/NE,2.5/2.5/SW,7.5/3.5/SE} {\edef\temp{%
	\noexpand\node[region,label] at(axis cs:\x,\y) {$\noexpand\sRegion{\l}$};
	
}\temp}
\addplot [sst,forget plot]
  table[row sep=crcr]{%
4 7\\4 4\\7 7\\4 7\\
};
\node[label] at (axis cs:5.5,6.5) {$\sSST$};
\node[below=1ex,label] at (axis cs:4,4) {$(\sPath[1]*,\sPath[1]*)$};
\node[above=2ex,label] at (axis cs:4,7) {$\sPath*=(\sPath[1]*,\sPath[2]*)$};
\node[right=1ex,label] at (axis cs:7,7) {$(\sPath[2]*,\sPath[2]*)$};
\end{axis}
\end{tikzpicture}
\endgroup

  }
  \caption{Domination relations for a c-t optimal point $\sPath*$ above the maximum $\sReprCCS{}$ ray: the arrows point to the greater factor value. The \sSST partitions $\sPIndSet$ in the $6$ marked areas.}\label{fig:ccs-domination}
\end{figure}

Combining the two domination lemmas \ref{thm:coverage-tendency-domination} and \ref{thm:representativeness-domination}, we can now prove the sufficiency of \tSST, by showing that every point outside the \tSST is dominated by one within $\sSST$. For this we distinguish two cases, depending on whether the c-t optimal point is above or below the maximum representativeness line.

Assume the optimal c-t point lies above the maximum $\sReprCCS{}$ ray. The point $\sPathOptInd$, along with the maximum representativeness ray, partition the \tCCS\ in the $6$ regions shown in Fig~\ref{fig:ccs-domination}, each of which has a non empty intersection with $\sSST$. We now show that the maxima of $\sffit$ over all the points in these regions, lie on this intersection, and therefore also in the \tSST.

We first study $\sRegion{SW}$: the points on the diagonal maximise $\sReprCCS{}$, while at the same time $\sCovCentCCS{}$ is dominated by the \tSST\ point $(\sPath[1]*,\sPath[1]*)$, therefore maximising $\sffit$ altogether. Similarly within $\sRegion{NE}$, we can show that $\sffit$ is maximised by $(\sPath[2]*,\sPath[2]*)\in\sSST$.

Within regions $\sRegion{W}$ and $\sRegion{N}$, both terms increase along each west-to-east and north-to-south path, respectively; these paths lead to a point of $\sSST$ that dominates all the rest on the traversed path. Within $\sRegion{SW}$, each west-to-east path ends up in a point of $\sRegion{N}$, which is itself dominated by a point of \tSST.

Finally, every south-to-north path of $\sRegion{SE}$ leads to either a point of $\sSST$ directly, or to one in the dominated $\sRegion{NE}$. We thus showed that no point of $\sPIndSet(\sSub)\setminus\sSST$ can maximise $\sffit$.

We work likewise if $\sPath*$ lies below the maximum $\sReprCCS{}$ ray.

\subsection{Proof of Proposition~\ref{thm:sequence-concavity}: concavity of $\sffitCCS{}$ along sequence}\label{sec:proof-sequence-concavity}

To prove this statement we employ the concavity of the sequences formed as the two factors $\sCovCentCCS{}$ and $\sReprCCS{}$ are computed along the horizontal and vertical sequences. We first treat the horizontal sequences, along which the entire $\sCovCentCCS{}\circ \sHSeq{}$ is concave, and so is $(\sReprCCS{}\circ \sHSeq{})(\siCnt)$ for the indices $\siCnt=0,\ldots,\siSub_2$, according to Lemmata~\ref{thm:optimal-ct-path} and \ref{thm:coverage-tendency-domination}, respectively.

Additionally, all factors are positive (or can be made so by adding an appropriate constant term) and so, raising the elements of the sequences to a power in $(0,1]$ preserves concavity. Multiplying the two re-weighted sequences yields
\begin{align}
\left((\sCovCentCCS{}\circ \sHSeq{})^\sweight(\sReprCCS{}\circ \sHSeq{})^{1-\sweight})\right)(\siCnt) = (\sffitCCS{}\circ\sHSeq{})(\siCnt)\;,
\end{align}
which is concave as the multiplication of two concave, positive sequences, therefore showing the concavity of the sequence of impact function values computed along the specified horizontal sub-sequence. Similarly we can work for vertical sequences.

Note that in our analysis we seamlessly interchange continuous and discrete convexity definitions. This is enabled by the uni-variate nature of the functions involved, since their discrete counterparts corresponds to sampling on regular intervals. Indeed, on one hand it can be shown that regular sampling of a uni-variate convex function yields a convex sequence \cite{yuceer_discrete_2002}. As a sufficiently applicable inverse for our needs, we can show that for any convex sequence, there exists at least one convex function with the same values at the sampled points and continuous second derivative; one such function results from cubic spline interpolation fitted on the sequence values.

\fi

\end{document}